\documentclass[lettersize,journal]{IEEEtran}
\usepackage{amsmath,amsfonts}
\usepackage{algorithmic}
\usepackage{algorithm}
\usepackage{array}
\usepackage[caption=false,font=footnotesize,labelfont=rm,textfont=rm]{subfig}

\usepackage{textcomp}
\usepackage{stfloats}
\usepackage{url}
\usepackage{verbatim}
\usepackage{graphicx}
\usepackage{cite}

\usepackage{bm}
\usepackage{booktabs}
\usepackage{epstopdf}
\usepackage{multirow}
\usepackage{float}
\graphicspath{{figures/}}

\newtheorem{theorem}{Theorem}
\newtheorem{definition}{Definition}

\begin{document}

\title{Game-Theoretic Machine Unlearning: Mitigating Extra Privacy Leakage}

\author{Hengzhu Liu, Tianqing Zhu, Lefeng Zhang, Ping Xiong$^*$
\thanks{Ping Xiong is the corresponding author. Hengzhu Liu and Ping Xiong are with the School of Information Engineering, Zhongnan University of Economics and Law, Wuhan, China (e-mail: hzliugrimm@stu.zuel.edu.cn; pingxiong@zuel.edu.cn).
Tianqing Zhu and Lefeng Zhang are with the Faculty of Data Science, City University of Macau, Macau (e-mail: tqzhu@cityu.edu.mo; lfzhang@cityu.edu.mo).}
\thanks{Manuscript received April 19, 2021; revised August 16, 2021.}}

\markboth{Journal of \LaTeX\ Class Files,~Vol.~14, No.~8, August~2021}%
{Shell \MakeLowercase{\textit{Liu et al.}}: A Sample Article Using IEEEtran.cls for IEEE Journals}


\maketitle

\begin{abstract}
With the extensive use of machine learning technologies,
data providers encounter increasing privacy risks.
Recent legislation,
such as GDPR,
obligates organizations to remove requested data and its influence from a trained model.
Machine unlearning is an emerging technique designed to enable machine learning models to erase users' private information.
Although several efficient machine unlearning schemes have been proposed,
these methods still have limitations.
First,
removing the contributions of partial data may lead to model performance degradation.
Second,
discrepancies between the original and generated  unlearned models can be exploited by attackers to obtain target sample's information,
resulting in additional privacy leakage risks.
To address above challenges,
we proposed a game-theoretic machine unlearning algorithm that simulates the competitive relationship between unlearning performance and privacy protection.
This algorithm comprises unlearning and privacy modules.
The unlearning module possesses a loss function composed of model distance and classification error,
which is used to derive the optimal strategy.
The privacy module aims to make it difficult for an attacker to infer membership information from the unlearned data,
thereby reducing the privacy leakage risk during the unlearning process.
Additionally,
the experimental results on real-world datasets demonstrate that 
this game-theoretic unlearning algorithm's effectiveness and its ability to generate an unlearned model with a performance similar to that of the retrained one while mitigating extra privacy leakage risks.
\end{abstract}

\begin{IEEEkeywords}
Machine unlearning,
membership inference attack,
game theory.
\end{IEEEkeywords}

\section{Introduction}
\IEEEPARstart{M}{achine} learning models are generally trained using large-scale data.
However,
the widespread application of these technologies has resulted in significant privacy threats for data providers.
As a result motivated legislation,
such as
the European Union’s General Data Protection Regulation (GDPR) 
and the California Consumer Privacy Act (CCPA), 
have legally established \emph{the right to be forgotten},
which provides users the authority to revoke personal data and its influence from a trained model.
To achieve the goal, a new technology emerged, named \emph{machine unlearning},
receiving considerable attention both in academia and industry \cite{xu2024,liu2024}.

A straightforward way to implement machine unlearning is removing the requested data
and retraining the model from scratch, 
but this type of methods results in high computational overhead \cite{cao2015}.
Current studies on machine unlearning mainly focus on designing efficient unlearning algorithms without retraining \cite{xu2024dont,xu2024update,Cha2024}.
Various machine unlearning methods have been proposed,
which can be categorized into two classes:
data-oriented unlearning and model-oriented unlearning.
Data-oriented techniques achieve unlearning through modifying the original training set.
For instance,
\emph{SISA} partitions the training data into multiple shards and trains submodels for each shard,
to effectively erase the unlearned data through retraining the corresponding submodels \cite{bourtoule2021}.
Model-oriented techniques involve manipulating the original model.
For example,
certified removal mechanism in \cite{guo2020} can eliminate the impact of specific samples through updating the model parameters.

While existing solutions have achieved significant performance in effective machine unlearning,
underlying problems of machine unlearning remain have not been addressed completely.
First,
to satisfy privacy protection requirements,
data removal conducted by machine unlearning usually reduces model performance \cite{graves2021}.
In order to improve unlearning efficiency,
some methods \cite{tarun2023deep} utilize a reference model,
trained on a subset of retain data,
to adjust and optimize the parameters.
This may weaken the model's performance due to a small amount of training samples being employed.
Second,
researchers have recently observed that machine unlearning generates additional privacy disclosure risks for the unlearned data \cite{chourasia2023}.
The differences between the initial and unlearned models,
such as prediction vectors and labels,
can be exploited by attackers to extract private information and infer the membership of target samples \cite{chen2021when,lu2022}.
These model discrepancies can also be adopted to reconstruct the unlearned data \cite{gao2022}.
Therefore,
establishing a trade-off between utility and privacy in machine unlearning is an urgent matter.
In this case, privacy refers to deleting particular data and its influence from a trained ML model,
and includes mitigating the potential privacy leakage risks during the post unlearning stage when providing inference services based on the unlearned model.


The fundamental cause of two above challenges is the over-unlearning. It means that the unlearning strategy has unlearned more information than we requested. If we consider the party who would steal information as an attacker while the one who unlearns the model as the defender, the most suitable unlearning strategy to tackle the above challenges might be finding a balance between the attacker and defender. One the one hand, the defender unlearns the exact information to specifically fit for the unlearning request, while on the other hand, the unlearned information cannot be too much so that the attacker can infer the information by comparing original and unlearned models. In this setting, 
we claim that if we consider the attackers and the defender of the machine unlearning procedure as a game, we can tackle the aforementioned challenges by leveraging the game theory.

Game theory is a theoretical framework to investigate the decision-making process of players,
where each participant aims to maximize payoff or minimize cost when selecting strategies \cite{kiennert2019}.
In machine unlearning, 
ensuring performance may increase additional privacy risks, 
while providing users with strong privacy protection can affect unlearning implementation. 
This implies that a competition between utility and privacy exists.
Thus,
game theory can be adopted to simulate interactive relationships and find stable solutions through equilibrium analysis.

In this paper,
we proposed a novel machine unlearning algorithm that takes advantage of game theory to model the unlearning process as a game, aiming to find the trade-off between utility and privacy.
Given the effectiveness of machine unlearning,
the attacker and defender are involved in this procedure as players,
Moreover,
we treated the attacker as a privacy evaluation mechanism,
and considered data removal and optimization of the ML model as game participants..
Specifically,
this unlearning algorithm consists of an unlearning and privacy module,
which sequentially select optimal strategies based on the loss minimization principle.
First,
the unlearning module's optimal strategy involves ensuring the efficacy of data removal.
We adopted an alternative retrained model to adjust the original model,
thereby providing an approximate statistical indistinguishability.
We also employed an error term to compensate for model performance degradation.
Second,
privacy module's strategy selection process is subject to the unlearning module's actions,
which holds an attack model to evaluate the privacy leakage risks of the unlearned model and update the parameters with the purpose of minimizing this risk,
thereby mitigating additional privacy threats.
After the game between the two modules,
the algorithm generates an unlearned model that maintains good performance and reduces privacy risks.
The contributions of this paper can be summarized as follows:
\begin{itemize}
	\item
    We investigated the trade-off between utility and privacy involved in machine unlearning,
    and proposed a novel game-theoretic unlearning algorithm.
	
	\item
    We mathematically quantified the privacy leakage risks caused by model discrepancy in machine unlearning
    and introduced privacy attack advantage as an evaluation metric.

    \item
    We provided theoretical analyses of the proposed unlearning algorithm,
    and proved the upper bound on the attacker's privacy advantage for the unlearned model.
    
	
\end{itemize}


\section{Related Work}\label{related work}

\subsection{Current machine unlearning solutions} \label{machine unlearning related work}

The concept of machine unlearning was first proposed by Cao et al. \cite{cao2015},
to erase the data and its impact on a trained machine learning model.
A naive way is to retrain from scratch on the retrained dataset;
however,
this incurs high computational costs.
Thus,
most existing studies focus on reducing the unlearning process overhead.
Some studies aim to implement data removal through manipulating the original training set \cite{bourtoule2021,ginart2019,tarun2021fast,chen2022graph,wang2023}.
For example,
Bourtoule et al. \cite{bourtoule2021} proposed an unlearning scheme,
\emph{SISA},
which divides the training set into several disjoint shards and independently trains an unique model on each shard.
When specific data need to be removed,
only the corresponding submodel requires retraining,
which can improve unlearning efficiency.
Tarun et al. \cite{tarun2021fast} developed an unlearning framework based on noise generation.
They utilized an error-maximizing noise matrix to disrupt the unlearned data's weights and restore performance using a repair function,
thereby retaining model accuracy.
Chen et al. \cite{chen2022graph} proposed a graph unlearning method,
\emph{GraphEraser},
which partitions a graph into several subgraphs that generate trained models to revoke data from the intermediate state.
Another type of unlearning methods directly modifies the original model's parameters to eliminate the data's impact \cite{guo2020,izzo2021,golatkar2020,golatkar2021,chundawat2023can,chundawat2023zero,wu2022}.
For instance,
Guo et al. \cite{guo2020} used influence theory \cite{koh2017} to quantify the data's impact,
and proposed a certified removal (CR) mechanism for approximate unlearning.
They erased specific training samples using a one-step Newton update \cite{koh2017} and adopted differential privacy loss perturbation to cover the gradient errors caused by removal.
Similarly,
Golatkar et al. \cite{golatkar2020} utilized the Newton update method to scrub a particular class based on a computable upper information bound value for the unlearned data \cite{martens2020} .
Chundawat et al. \cite{chundawat2023can,chundawat2023zero}studied unlearning in deep neural networks (DNNs).
They developed an unlearning method based on a student-teacher framework, 
which effectively eliminates the unlearned data's influence \cite{chundawat2023can}.
They also proposed a zero-shot unlearning approach, based on gated knowledge transfer, to remove sample information while maintaining the model performance on the retain set \cite{chundawat2023zero}.

\subsection{Privacy risks in machine unlearning}

Several privacy attacks against machine learning have been proposed,
such as membership inference attacks \cite{shokri2017} and attribute inference attacks \cite{liu2022mldoctor}.
These attacks aim to steal sensitive data and model information. 
For example,
membership inference attacks can obtain data membership information,
determining whether a particular sample belongs to the training set.
As mentioned in Section \ref{machine unlearning related work},
there are many machine unlearning methods designed to satisfy the privacy protection requirements.
However,
recent studies have demonstrated that machine unlearning also introduces additional privacy risks \cite{chen2021when,stock2023,gao2022,chourasia2023}.
Machine unlearning techniques modify the original model to achieve data removal,
resulting in an unlearned model.
The discrepancy between the original and unlearned model,
such as model parameters and posteriors,
which may contain the private information about unlearned data \cite{baumhauer2022}.
Chen et al. \cite{chen2021when} investigated the privacy issues caused by machine unlearning.
They proposed a novel membership inference attack strategy against machine unlearning,
which employs the posterior differences generated by the original and unlearned models for the same samples to construct features to acquire the unlearned data's membership,
providing evidence for privacy leakage caused by unlearning.
Similarly,
Gao et al. \cite{gao2022} presented a deletion inference and reconstruction framework in machine unlearning,
which successfully extracts information about the unlearned data.
Additionally,
potential defense strategies for solving privacy issues also exist.
For example,
differential privacy (DP) has been proven to be effective in limiting the influence of samples on outputs \cite{jayaraman2019,liu2022mldoctor}.
Therefore,
using DP to train models can alleviate membership inference attacks against unlearning \cite{chen2021when};
however, it compromises the model performance.
Jia et al. \cite{jia2019} developed a defense scheme that adds noise to the confidence vectors to weaken membership inference attacks.



\section{Preliminaries} \label{preliminaries}

\subsection{Notations}

Let $D=\{(\bm{{\rm x_1}}, y_1), (\bm{{\rm x_2}}, y_2), \ldots, (\bm{{\rm x_i}}, y_i)\}$ denote the dataset,
consisting of the training set $D_{train}$ and the test set $D_{test}$.
The original model
$M_o$ is trained on $D_{train}$ by a machine learning algorithm $\mathcal{A}$,
namely,
$M_o = \mathcal{A}(D_{train})$.
$D_{test}$ is used for model evaluation.
In machine unlearning,
given the unlearned data $D_f$,
$D_r$ represents the retain training set with $D_f$ deleted (i.e., $D_r = D_{train} \setminus D_f$).
One way to unlearning is retraining the model on $D_r$ to obtain $M_r = \mathcal{A}(D_r)$,
while an unlearning algorithm $\mathcal{U}$ acquires the unlearned model $M_u$,
which should be statistically indistinguishable from $M_r$.
Specific to our method,
we introduce in our framework,
an unlearning module and a privacy module.
As retraining is costly,
the unlearning module targets the alternative $M_r^{'}$ trained on $D_r^{'}$ ($D_r^{'} \subseteq D_r$) to tune the original model $M_o$.
It also utilizes $D_{third}$,
which is a subset of the test set, to compensate for the performance loss.
The privacy module measures the privacy risks using a membership attack model $M_A$.


\subsection{Machine unlearning} \label{mul-definition}

We expect to find an unlearning algorithm $\mathcal{U}$,
which acquires the unlearned model $M_u$ that performs similarly to the retrained model $M_r$ without retraining.
Specifically,
for the same sample, $M_o$ and $M_r$ should have consistent or statistically indistinguishable output distributions.
In other words,
the distribution of the weights of $M_u$ and $M_r$ should be identical or statistically indistinguishable.
The formal definition of machine unlearning is as follows:

\begin{definition}[Machine Unlearning \cite{bourtoule2021}]
	Given a machine learning algorithm $\mathcal{A}(\cdot)$,
	a training set $D_{train}$ and the unlearned data $D_f$.
	$M_o$ is the original model trained on $D_{train}$ using $\mathcal{A}(\cdot)$.
	For an unlearning process $\mathcal{U}(\cdot)$,
	we say $D_f$ is unlearned from $M_o$ if:
	\begin{equation}\label{mulequation}
		\mathcal{P}(\mathcal{U}(\mathcal{A}(D_{train}), D_{train}, D_f)) \cong  \mathcal{P}(\mathcal{A}(D_{train}\setminus D_f))
	\end{equation}
	where $\mathcal{P}(\cdot)$ represents the weight distribution.
\end{definition}

\subsection{Membership inference attacks}

Membership inference attacks are a typical method that steals private information of training data,
aiming to infer whether a sample is a member of a target model's training set.

\begin{definition}
	[Membership Inference Attacks \cite{shokri2017}]
	Given a machine learning model $M_o$,
	a training set $D_{train}$ and a target sample $\bm{{\rm x_i}}$.
	Let I($\cdot$)  be an indicator function for membership inference attacks,
	used to determine the membership information of $\bm{{\rm x_i}}$.
	\begin{equation}
		I(\bm{{\rm x_i}}) = 
		\begin{cases} 
			1 & \text{if } \bm{{\rm x_i}}\text{ is a member of } D_{train} \\
			0 & \text{otherwise}
		\end{cases}
	\end{equation}
	
	$I(\bm{{\rm x_i}}) = 1$ indicates that $\bm{{\rm x_i}}$ is a member of $D_{train}$,
	otherwise $I(\bm{{\rm x_i}}) = 0$.
\end{definition}


Unlike traditional membership inference attacks that only utilize information from the target model,
attacks against machine unlearning \cite{chen2021when} employ the combined features of $M_o$ and $M_u$ generated by unlearning to train the attack model $M_A$.
The attacker queries $M_o$ and $M_u$ to obtain the corresponding posteriors and constructs the feature vector $\mathcal{F}$ to train the attack model $M_A$.
Then,
sending the posteriors of these two models for $\bm{{\rm x_i}}$,
$\mathcal{F}(\bm{{\rm x_i}}) = (M_o(\bm{{\rm x_i}}), M_u(\bm{{\rm x_i}}))$ to $M_A$,
and its output,

\begin{equation}
	M_A(M_o(\bm{{\rm x_i}}), M_u(\bm{{\rm x_i}}))
\end{equation}
is the probability that $\bm{{\rm x_i}}$ belongs to the training set of $M_o$, but it is not in the training dataset of $M_u$.

\subsection{Game theory}


Stackelberg game is a classical game theory,
involving two decision makers (i.e., players):
a leader and follower.
It is a sequential game where the leader makes decisions first,
and the follower responds after observing the leader's actions.
Let $L$ be the leader,
and $F$ represents the follower.
In Stackelberg game,
each player has a strategy space,
where $L$'s strategy space is denoted as $\mathcal{S}_L =  \{s_L^1, s_L^2, \ldots, s_L^i\}$ (i.e., $s_L^i$ is the $i$-th $L$'s strategy),
and $F$'s strategy space is $\mathcal{S}_F = \{s_F^1, s_F^2, \ldots, s_F^j\}$ ($s_F^j$ is the $j$-th $F$'s strategy).
For strategy selection,
each player also has a payoff function,
and the optimal strategy maximizes the payoff or minimizes the cost.
$R_L$ and $R_F$ represent payoff functions of leader and follower,
respectively.

In every game,
the values of $R_L$ and $R_F$ depend on $L$ and $F$'s candidate strategies, $s_L^i$ and $s_F^j$, respectively.
The leader's objective is to find an optimal strategy with the best payoff,
as shown in Eq. \ref{Lpayoff}:

\begin{equation} \label{Lpayoff}
	s_L^{*}=\operatorname*{argmax} _{s_L \in \mathcal{S}_L} R_L\left(s_L, s_F^{*}\right)
\end{equation}

In response to the strategy chosen by $L$,
the best action for $F$ is to maximize payoff based on $s_L^{*}$:

\begin{equation} \label{Fpayoff}
	s_F^{*}=\operatorname*{argmax} _{s_F \in \mathcal{S}_F} R_F\left(s_L^{*}, s_F\right)
\end{equation}

Combining Eq. \ref{Lpayoff} and Eq. \ref{Fpayoff} results in Eq. \ref{LFpayoff},
which defines the Nash equilibrium under the condition of final strategies $s_L^{*}$ and $s_F^{*}$ after the game.

\begin{equation} \label{LFpayoff}
	\left(s_L^{*}, s_F^{*}\right)=\operatorname*{argmax}_{s_L \in \mathcal{S}_L} R_{L}
	\left(s_L, \operatorname*{argmax}_{s_F \in \mathcal{S}_F} R_{F}(s_L, s_F)\right)
\end{equation}

\section{Game-theoretic machine unlearning method} \label{method}

\subsection{Problem definition}

In this paper,
we focus on the classification tasks in machine learning,
including binary and multi-class classification.
Machine learning models typically take the users' data (i.e. training set $D_{train}$) as input
and obtain the original model $M_o$ using the machine learning algorithm $\mathcal{A}(\cdot)$.
After the model is trained,
the user submits an unlearning request to revoke data $D_f$,
and the model provider erases the influence of $D_f$ from $M_o$ without the costly retraining.
Thus,
The goal of this study is to develop an unlearning algorithm $\mathcal{U}(\cdot)$ to find an unlearned model $M_u$ which performs similarly to the retrained one while providing the privacy guarantee.

\textbf{Threat model:} There is an attacker who aims to acquire the target sample's membership information,
namely whether it was used to train the original model $M_o$ and is unlearned from $M_o$.
In our setting,
the attacker is assumed to have black-box access to the original model $M_o$ and the unlearned model $M_u$.
Specifically,
the attacker only has the capability to query the model and obtain the corresponding prediction vectors,
and the target sample's membership can be determined through implementing membership inference attacks.

The defender is the model provider who possesses a classifier and an unlearning algorithm,
and the goal is to ensure the effectiveness of unlearning and lower the privacy leakage probability caused by the membership inference attack.
Specifically,
the defender can query the model in a white-box manner,
and has knowledge of model parameters and training data.
Moreover,
the unlearning algorithm $\mathcal{U}(\cdot)$ generates an unlearned model denoted as $M_u$,
which is not only statistically indistinguishable from the retrained model $M_r$,
but also make it difficult for the attacker to determine the membership information of $D_f$ based on the outputs of $M_o$ and $M_u$.
In this paper,
we define the attacker's ability to infer membership information as privacy attack advantage,
which refers to the difference between the attacker's inference probability and the parameter $\lambda$.
For the defender,
the goal of reducing privacy risks can be formalized as diminishing the privacy attack advantage.
Mathematically, 
the unlearning problem is formulated as follows:

\begin{equation}
	\begin{cases} 
		\mathcal{P}(M_u(D_{train}\setminus D_f) \cong 
		\mathcal{P}(M_r(D_{train}\setminus D_f) \\
		M_A(M_o(D_f), M_u(D_f)) \rightarrow \lambda
	\end{cases}
	\label{problem}
\end{equation}
where $\lambda$ is a constant between $(0,1)$.
Ideally,
$\lambda$ should be equal to $0.5$,
signifying that the probability of $D_f$ being inferred by the attacker as a member of the original training set is close to $0.5$.
In the subsequent theoretical analysis and main experiments,
we set $\lambda$ to $0.5$ as this is the most ideal scenario.

The trade-off between utility and privacy is indeed the primary issue that needs to be addressed in the design of unlearning algorithms. 
Thus, in this paper,
the defender possesses an unlearning module and a privacy module,
and the unlearning process is implemented through the game between these two modules.
The resulting unlearned model $M_u$ is expected to be similar to the retrained model with a reduced privacy leakage risk.

\subsection{The overview of proposed method}

\begin{figure*}[h]
	\centering
	\includegraphics[scale=0.45]{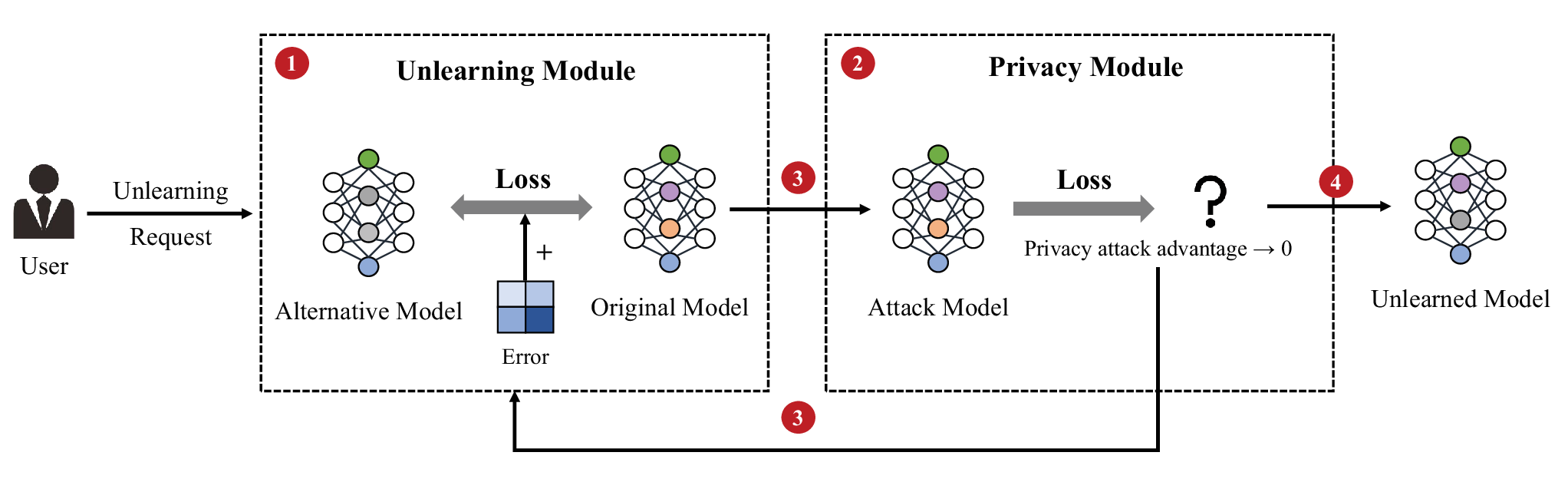}
	\caption{The overview of proposed method}
	\label{mulmethod}
\end{figure*}

The game-theoretic machine unlearning algorithm consists of two components:
the unlearning module and the privacy module,
as depicted in Fig. \ref{mulmethod}.
The unlearning module aims to ensure the effectiveness of machine unlearning and maintain model performance,
while the privacy module mitigates the privacy risks arising from membership inference attacks.

As shown in Fig. \ref{mulmethod},
the proposed method involves a sequential Stackelberg game process with the unlearning and privacy modules as the players.

\begin{itemize}
	\item 
	First,
	the model provider trains the original model $M_o$ on the training set $D_{train}$.
    Then,
	a user submits an unlearning request to withdraw the use of data $D_f$.
	The game-theoretic unlearning algorithm begins to erase $D_f$ and its influence on $M_o$.
	
	\item
	Second,
	as the leader of the game,
	the unlearning module initially moves to find the optimal strategy for adjusting the parameter $\omega_o$ of $M_o$.
	We train an alternative model $M_r^{'}$ to be retrained on a subset of the retain set $D_r$,
    to calculate the distance between $M_o$ and $M_r^{'}$.
	Meanwhile,
	a classification error is added to the loss to maintain model performance on $D_r$.
    Thus,
    the goal of unlearning module is to minimize the loss comprising the distance and the error term.
    
	\item 
	Next,
	as the follower,
	the privacy module's action to mitigate privacy risks,
	is subject to the unlearning module's strategy,
	namely,
	the parameter $\omega_u$ of the output $M_u$.
	This module possesses an attack model,
	and optimizes $\omega_u$ through minimizing the privacy advantage of the attacker,
	making it difficult to infer the target samples' membership information.
	
	\item 
	After that,
	the unlearning and privacy modules repeat the process to modify $\omega_u$ based on the corresponding loss,
	and after all the game play,
	the modules converge to an equilibrium state.
	
	\item 
	Finally,
	at equilibrium, 
	the parameter $\omega_u$ is the unlearned model we are finding,
	which satisfies the requirements for privacy protection while achieving performance similar to the model retrained from scratch.
	
\end{itemize}

\subsection{The unlearning module}

The unlearning module aims to erase the unlearned data $D_f$ and its influence from the trained original model $M_o$ to ensure the effectiveness of machine unlearning.
Specifically,
it aims to find an unlearned model $M_u$ that performs similarly to the retrained model $M_r$ based on the definition described in Section \ref{mul-definition}.
In the established Stackelberg game framework,
the unlearning module is regarded as the leader $L$ of game,
which initially takes action.

The unlearning module with a strategy space $\mathcal{W} = \{\omega_1, \omega_2, \ldots, \omega_i \}$,
which includes all potential model parameters $\omega$ of the unlearned model $M_u$,
selects an optimal strategy, $\omega^{*}_L$ that minimizes the loss.
Specifically,
$L$'s primary task is to satisfy the core machine unlearning requirements and
remove $D_f$ from the original model $M_o$.
As shown in Eq. \ref{mulequation},
for the same input sample,
the output distributions of the unlearned model $M_u$ and retrained model $M_r$ are desired to be statistically indistinguishable.
Nevertheless,
we recognize that retraining is time-consuming,
and in real-world scenarios,
obtaining $M_r$ before the unlearning request arrives is impossible.
Therefore,
inspired by \cite{chen2021lightweigh,tarun2023deep},
we employ an alternative model $M_r^{'}$ to $M_r$ as the optimization direction for the unlearning process.
This alternative model $M_r^{'}$ is trained on a subset comprising a few samples from retain training set $D_r^{'}$.
Additionally,
we also select the third-party data $D_{third}$,
which is independent of $D_{train}$ ($D_{third} \cap D_{train} = \emptyset$),
to examine the similarity between $M_u$ and $M_r^{'}$.
This process can be formalized as:

\begin{equation} \label{unlearn-third}
	\mathcal{P}(M_u(D_{third})) \cong \mathcal{P}(M_r^{'}(D_{third})) 
\end{equation}

As mentioned above,
the alternative model $M_r^{'}$ is trained on a small amount of data,
so that its performance is naturally weaker than that of the retrained model $M_r$.
This results in using $M_r^{'}$ as the unlearning operation reference impairs $M_u$'s performance.
Consequently,
we add an error term $error(M_u)$ to the unlearning module to compensate for the performance loss, as expressed in Eq. \ref{error}.

\begin{equation} \label{error}
	error(M_u) = -\sum_{c=1}^{C} y_{c} \log(\hat{y}_c)
\end{equation}
where $y_c$ is the target sample's true label,
and $\hat{y}_c$ is the probability distribution of the model $M_u$ predicting the $c$-th class.

Combing Eq. \ref{unlearn-third} and Eq. \ref{error} results in the loss function $Loss_L$ of the unlearning module in Eq. \ref{loss-l}.
Additionally,
the distance between $M_u$ and $M_r^{'}$ evaluates the unlearning process's effectiveness.

\begin{align} \label{loss-l}
	Loss_L & = Dis(M_u(D_{third}), M_r^{'}(D_{third})) + error(M_u) \notag \\
	& = \sqrt{\sum_{i=1}^{n} (M_u(D_{third}) -M_r^{'}(D_{third}))^2} -\sum_{c=1}^{C} y_{c} \log(\hat{y}_c)
\end{align}

Consequently,
in each game round,
the optimal strategy $\omega^{*}_L$ of the unlearning module ($L$) minimizes the loss function $Loss_L$,
which can also be written as:

\begin{align} \label{unlearning-l}
	& \operatorname*{argmin} Loss_L  \notag \\
	& = \operatorname*{argmin} Dis(M_u(D_{third}), M_r^{'}(D_{third})) + error(M_u) 
\end{align}

The unlearning module's complete algorithm is presented in Algorithm \ref{unlearning-a}.
The algorithm takes as input the parameters of the original model $\omega_o$,
the alternative model $M_r^{'}$,
the third-party data $D_{third}$, and the retain training set $D_r$,
adjusting the original model to acquire the unlearned one $M_u$ after data removal.
The algorithm initializes the parameters of $M_u$ to $\omega_o$ in the first iteration (line 1).
For each epoch,
it first computes the Euclidean distance between the output distributions of the unlearned model $M_u$ and the alternative model $M_r^{'}$ on each $D_{third}$ sample.
Next,
the classification error (i.e., the cross-entropy error of $M_u$ on the retain set $D_r$) is calculated to recover model performance (line 2-4).
Finally,
the algorithm constructs a loss function,
and updates the parameters of $M_u$ to erase the influence of $D_f$ using gradient descent,
where $\eta$ is the learning rate (line 5-7).

\begin{algorithm}[h]
	\caption{The Algorithm in Unlearning Module}
	\label{unlearning-a}
	\begin{algorithmic}[1]
		\REQUIRE the initial parameters $\omega_o$ trained by machine learning $\mathcal{A}(\cdot)$,
		the alternative model $M_r^{'}$ trained on the subset of training set,
		the third-party data $D_{third}$,
		the retain training set $D_r$.
		\ENSURE the unlearned model parameters $\omega_u$
		
		\STATE Initialize the unlearned model as $\omega_o$.
		\FOR{epoch}
		\STATE \textbf{compute} the distance between $M_u$ and $M_r^{'}$.
		
		\begin{equation*} 
			Dis(M_u, M_r^{'}) = \sqrt{\sum_{i=1}^{n} (M_u(D_{third}) -M_r^{'}(D_{third}))^2}
		\end{equation*}
		where $n$ is the vector dimension.
		
		\STATE \textbf{compute} the classification error of $M_u$ on $D_r$.
		
		\begin{equation*} 
			error(M_u) =  -\sum_{c=1}^{C} y_{c} \log(\hat{y}_c)
		\end{equation*}
		
		\STATE \textbf{update} the unlearned model $M_u$ by
		
		\begin{align*} 
			& \omega_u = \omega_u - \eta \nabla_{\omega_u} Loss_L \\
			& s.t.  Loss_L = Dis(M_u, M_r^{'}) + error(M_u)
		\end{align*}
		where $\eta$ is the learning rate.
		
		\ENDFOR
		\RETURN $\omega_u$
	\end{algorithmic}
\end{algorithm}

\subsection{The privacy module}

The privacy module aims to mitigate the privacy risks caused by the discrepancy between the original model $M_o$ and the unlearned model $M_u$.
Specifically,
this module adjusts the parameters $\omega_u$ of $M_u$ generated by the unlearning module,
making it difficult for the attacker to distinguish between the outputs of $M_o$ and $M_u$ for the target samples.
In the Stackelberg game,
we define the privacy module as the follower $F$,
which responds after observing the leader's moves.

Similarly,
the privacy module also has a strategy space $\mathcal{W} = \{\omega_1, \omega_2, \ldots, \omega_i \}$ shared with $L$,
to find the best action $\omega_F^{*}$ to reduce privacy leakage risks.
Specifically,
$F$ holds a membership inference attack model $M_A$,
which is trained on the comprehensive features of $M_o$ and $M_u$ based on the method proposed by Chen et al. \cite{chen2021when}.
The output of $M_A$ is a probability value $M_A(M_o(D_f), M_u(D_f))$,
representing the probability that the unlearned data $D_f$ belongs to the original training set $D_{train}$ of $M_o$.
According to the problem definition shown in Eq. \ref{problem},
the ideal attack probability is expected to approximate the random guessing level (i.e., $\lambda = 0.5$),
and we define the difference between this probability and $0.5$ as the privacy attack advantage $PriAA$:

\begin{equation} \label{priaa}
	PriAA = \left\lvert M_A(M_o(D_f), M_u(D_f)) - 0.5 \right\rvert 
\end{equation}

Thus,
the privacy module utilizes $PriAA$ as the loss function $Loss_F$ to obtain the optimal strategy.
Mathematically,
given the optimal strategy $\omega^{*}_L$ already selected by $L$,
$F$'s best response $\omega^{*}_F$ should be able to minimize $Loss_F$ as shown in Eq. \ref{optimal-f}.

\begin{align} \label{optimal-f}
	\operatorname*{argmin} Loss_F
	= & \operatorname*{argmin} \left\lvert M_A(M_o(D_f), M_u(D_f)) - 0.5 \right\rvert 
\end{align}




Algorithm \ref{optimization-a} displays the privacy module's strategy selection process.
The algorithm takes the adjusted unlearning model's parameters $\omega_u$ from the unlearning module, 
the original model $M_o$, 
the unlearned data $D_f$, and the well-trained member inference attack model $M_A$ as inputs,
aiming to reduce the privacy risks caused by unlearning.
For each epoch,
this algorithm starts by querying the original model $M_o$ and the unlearned model $M_u$,
respectively,
to obtain the confidence vectors for the unlearned data $D_f$ (line 2-3).
Then,
it employs these confidence vectors to calculate the attack probability and privacy attack advantage of $M_A$ (line 4).
Finally,
the algorithm updates $M_u$ by minimizing the privacy attack advantage,
which is formalized as the absolute difference between the membership probability and $0.5$,
to protect the private information of $D_f$.

\begin{algorithm}[h]
	\caption{The Algorithm in Privacy Module}
	\label{optimization-a}
	\begin{algorithmic}[1]
		\REQUIRE the model parameters $\omega_u$ adjusted by the unlearning module,
		the parameters $\omega_o$ of the original model $M_o$,
		the membership inference attack model $M_A$,
		the unlearned data $D_f$.
		\ENSURE the unlearned model parameters $\omega_u$
		
		\STATE Initialize the unlearned model as $\omega_u$ from the unlearning module.
		\FOR{epoch}
		\STATE \textbf{compute} the outputs of $M_o$ and $M_u$ for $D_f$, respectively.
		\STATE \textbf{compute} the privacy attack advantage of attacker:
		
		\begin{equation*} 
			PriAA = \left\lvert M_A(M_o(D_f), M_u(D_f)) - 0.5 \right\rvert 
		\end{equation*}

		\STATE \textbf{update} the unlearned model $M_u$ by
		
		\begin{align*} 
			& \omega_u = \omega_u - \eta \nabla_{\omega_u} Loss_F \\
			& s.t. Loss_F = \left\lvert M_A(M_o(D_f), M_u(D_f)) - 0.5 \right\rvert
		\end{align*}
		
		\ENDFOR
		\RETURN $\omega_u$
	\end{algorithmic}
\end{algorithm}

\subsection{Analysis of the game}

We present a theoretical analysis of the game-theoretic properties for the proposed unlearning algorithm.
Specifically,
we analyze the existence of an equilibrium and the convergence of the game by combing the differentiable properties of Algorithm \ref{unlearning-a} and Algorithm \ref{optimization-a}.

According to the loss functions of the unlearning and privacy modules,
we deduce that in every game play,
the optimal strategy $\omega_L^{*} \in \mathcal{W}$ of the leader (i.e., unlearning module) minimizes the loss $Loss_L$:

\begin{align} \label{best-l}
	\omega_L^{*} & = \operatorname*{argmin}_{\omega_L \in \mathcal{W}} Loss_L(\omega_L, \omega_F) \notag \\
	& = \operatorname*{argmin}_{\omega_L \in \mathcal{W}} Dis(M_u, M_r^{'}) + error(M_u) 
\end{align}

In response to the privacy risks of $\omega_L^{*}$,
the best move of the follower (i.e., privacy module) converges onto a $\omega_F^{*} \in \mathcal{W}$ with the lowest $Loss_F$:

\begin{align} \label{best-f}
	\omega_F^{*} & = \operatorname*{argmin}_{\omega_F \in \mathcal{W}} Loss_F(\omega_L, \omega_F) \notag \\
	& = \operatorname*{argmin}_{\omega_F \in \mathcal{W}} \left\lvert M_A(M_o(D_f), M_u(D_f)) - 0.5 \right\rvert
\end{align}

Substituting Eq. \ref{best-f} in Eq. \ref{best-l},
we derive the following theorem:

\begin{theorem} \label{nashequilibrium}
	The unlearning module and the privacy module sequentially search for optimal strategies,
	finally converging to an equilibrium, as shown in Eq. \ref{nash}.
	\begin{align} \label{nash}
		& \left(\omega^{*}_L, \omega^{*}_F\right)=\operatorname*{argmin}_{\omega_L \in \mathcal{W}} Loss_L 
		\left(\omega_L, \omega^{*}_F\right) \notag \\
		& s.t. \quad \omega^{*}_F = \operatorname*{argmin}_{\omega_F \in \mathcal{W}} Loss_F(\omega_L, \omega_F)
	\end{align}
\end{theorem}

Theorem \ref{nashequilibrium} characterizes the conditions where the game-theoretic method achieves an equilibria.
At this point,
the leader's optimal strategy $\omega^{*}_L$ is selected based on the follower's best response $\omega^{*}_F$,
while $\omega^{*}_F$ is selected knowing the leader's move.
The equilibrium is the intersection of these two choices,
which means that neither player would want to unilaterally change their strategy.
Note that although each player at the equilibrium holds an optimal strategy,
these moves are similar in our method, with $\omega^{*}_F$ being the final obtained unlearned model parameters.
We prove Theorem \ref{nashequilibrium} below.

\emph{Proof of Theorem \ref{nashequilibrium}:}
To demonstrate Theorem \ref{nashequilibrium}, 
the convergence of Algorithm \ref{unlearning-a} and Algorithm \ref{optimization-a} needs to be proven.
For $L$,
the gradient descent is employed to perform a stochastic search for the optimal strategy $\omega_L^{*}$,
which minimizes the loss function $Loss_L$.
This process can be expressed as:

\begin{align} \label{g-l}
	& \omega_L^{*} = \omega_u - \eta \nabla Loss_L, \notag \\
	& s.t. \quad Loss_L =  \sqrt{\sum_{i=1}^{n} (M_u(D_{third}) -M_r^{'}(D_{third}))^2}  \notag \\
	&  + (-\sum_{c=1}^{C} y_{c} \log(\hat{y}_c))
\end{align}
where $\eta$ is the learning rate,
and $\nabla$ is the gradient of $Loss_L$ with respect to the parameter $\omega_u$.
In the first round of the game, 
the $\omega_u$ value is the parameters $\omega_o$ of the original model.
In the subsequent rounds, 
$\omega_u$ represents the intermediate values.

Using the first-order Taylor’s expansion, 
we obtain:
\begin{equation}
	Loss_L\left(\omega_L^{*}\right) \approx Loss_L\left(\omega_u \right)+\nabla Loss_L\left(\omega_u \right)^{T}\left(\omega_L^{*}-\omega_u \right)
\end{equation}

Substituting Eq. \ref{g-l}:
\begin{equation}
	Loss_L\left(\omega_L^{*}\right) \approx Loss_L\left(\omega_u \right)-\eta\left\|\nabla Loss_L\right\|^{2}
\end{equation}

Let $\bm{{\rm x}}$ be a sample of $D_{third}$.
Thus,
$\mathbb{E}[p(\bm{{\rm x}})]$ and $\mathbb{E}[r(\bm{{\rm x}})]$ are the mean vectors of $M_u(\bm{{\rm x}})$ and $M_r^{'}(\bm{{\rm x}})$,
respectively.
$\nabla Loss_L$ can be formalized as:

\begin{equation} \label{nabla-l}
	\nabla Loss_L = 2(\mathbb{E}[p(\bm{{\rm x}})]-\mathbb{E}[r(\bm{{\rm x}})]) \cdot \nabla \mathbb{E}[p(\bm{{\rm x}})]-\sum_{c=1}^{C} y_{c} \frac{\partial \log \left(\hat{y}_c(\bm{{\rm x}})\right)}{\partial \bm{{\rm x}}}
\end{equation}

Both terms on the right-hand side of Eq. \ref{nabla-l} are differentiable.
Therefore,
according to the chain rule,
the combination $\nabla Loss_L$ is also differentiable.
Since $\left\|\nabla Loss_L\right\|^{2}$ is non-negative,
and $\eta$ is a positive learning rate,
we can obtain $Loss_L\left(\omega_L^{*}\right) \leq  Loss_L\left(\omega_u \right)$.
Thus,
the gradient descent in the unlearning module is convergent for searching the optimal strategy.

%
%

Similarly,
in Algorithm \ref{optimization-a},
the follower $F$ searches for the best response $\omega^{*}_F$ under the action of the leader $\omega^{*}_L$ through gradient descent:

\begin{align} \label{g-f}
	& \omega_F^{*} = \omega_u - \eta \nabla Loss_F, \notag \\
	& s.t. \quad Loss_F = \left\lvert M_A(M_o(D_f), M_u(D_f)) - 0.5 \right\rvert
\end{align}
where the output of $M_A$ is a probability value obtained through a differentiable softmax activation function,
indicating that $\nabla Loss_F$ is differentiable.
Based on the above proof results, 
we deduce the following:
\begin{equation}
	Loss_F\left(\omega_F^{*}\right) \approx Loss_F\left(\omega_u \right)-\eta\left\|\nabla Loss_F\right\|^{2}
\end{equation}
That indicates that the strategy selection of $F$ is also convergent.
Therefore,
during the game process, 
the unlearning and privacy modules successively adjust $\omega_u$,
which is a locally optimal solution found using a convergent gradient descent rule. 
Ultimately,
the players' optimal strategies tend towards achieving stability.
That completes the proof.

\subsection{Privacy analysis}

In this section,
we theoretically analyze the effectiveness of our method in mitigating privacy risks. 
The privacy guarantee is given by the following theorem:

\begin{theorem}\label{privacy-theorem}
	There exists $k(k > 0)$ that makes the probability of membership inference attack against unlearning approaches $0.5$.
	Specifically,
	\begin{equation} 
		\label{privacy-bound}
		0.5-k \leq P(M_A(M_o(D_f), M_u(D_f))) \leq 0.5+k
	\end{equation}
\end{theorem}

Theorem \ref{privacy-theorem} illustrates the bounds of privacy risk,
enabling the proposed method to obtain the optimal game strategy.

\emph{Proof of Theorem \ref{privacy-theorem}:}
The confidence vector output by the membership inference attack model $M_A$ is denoted as $X = P(M_A(M_o(D_f), M_u(D_f)))$,
representing the probability that target sample (i.e., $D_f$) belongs to the original training set.
$X$ is a continuous random variable with variance $\sigma^2$ and expectation $\mu$.
Theorem \ref{privacy-theorem} states that it is difficult for the attacker to extract additional information from the original and unlearned models' outputs; therefore,  the attack has no advantage.
Thus,
$X$ is expected to be close to the random guessing's probability $0.5$ (i.e., $\mu$ equals to $0.5$).

According to Eq. \ref{privacy-bound},
we set $\mu = 0.5$.
By applying the Chebyshev inequality,
we derive the following:
\begin{align} 
	P\{|X-0.5| \geq k\} & = \int\limits_{|X-0.5| \geq k}f(x)dx \notag \\
	& \leq \int\limits_{|X-0.5| \geq k}\frac{(X - 0.5)^{2}}{k^2} f(x)dx  \notag \\
	& \leq \frac{1}{k^2} \int_{- \infty}^{+ \infty} (X - 0.5)^{2} f(x)dx = \frac{\sigma^2}{k^2}
\end{align}

That is
\begin{equation}
	\label{bound}
	P\{|X-0.5| \geq k\} \leq \frac{\sigma^2}{k^2}
\end{equation}
where $k$ is a positive constant and $\frac{\sigma^2}{k^2}$ is the upper bound of privacy risks.
In Eq. \ref{bound},
the larger the value of $k$,
the smaller the upper bound,
indicating that $X$ does not deviate significantly from $0.5$.
Specifically, 
the probability of $X$ falling around $0.5$ is very high.
According to Algorithm \ref{optimization-a},
we know that the privacy module aims to find the optimal $k$ in every game play,
and adjusts the parameters $\omega_u$ under the upper bound's constraints. That completes the proof.

\subsection{Case study}

In this subsection,
we introduce a realistic case to describe the specific process of the proposed game-theoretic unlearning algorithm.
Suppose a company focused on medical image analysis,
collecting a large amount of medical image data from patients,
denoted as $D_{train}$,
to train a machine learning model $M_o$ that has the capability to identify critical features in medical images to assist in disease diagnosis.
Due to increased privacy protection awareness,
a patient $u$
submits an unlearning request to this institution to revoke their data $D_f$ from the trained model.
The company begins the unlearning process to eliminate the influence of $D_f$ from $M_o$,
which is accomplished through the game between the unlearning and privacy modules.
Specifically,
the game consists of the following steps:

\begin{itemize}
	\item 
	\textbf{Step 1: The unlearning module's strategy selection.} 
	As the leader of game,
	the unlearning module initially seeks the optimal strategy $\omega_L^{*}$ to modify the parameters $\omega_o$ of $M_o$,
	aiming to erase the influence of $D_f$ on the model.
	The company trains an alternative model to tune $\omega_o$, 
	recovers model performance using classification errors,
	and minimizes loss according to Eq. \ref{unlearning-l}, obtaining the optimal strategy $\omega_L^{*}$.
	
	\item 
	\textbf{Step 2: The privacy module's strategy selection.}
	As the follower,
	the privacy module responds to the leader's optimal strategy $\omega_L^{*}$.
	Specifically,
	actions depend on $\omega_L^{*}$,
	and it uses an attack model to measure the privacy risk of $M_u$,
	and minimizes the advantage of the attack model according to Eq. \ref{optimal-f}, employing the optimal strategy $\omega_F^{*}$ to adjust $M_u$.
	
	
	\item 
	\textbf{Step 3: Iterative modification and equilibrium solution.}
	After repeating Steps 1 and 2, 
    the unlearning and privacy modules iteratively modify $\omega_u$,
	which is subjected to the strategy selected by the other player.
    Then, compute the local optimal solution based on the gradient descent principle according to the corresponding loss function.
	
	
\end{itemize}

Ultimately,
after $100$ iterations,
the losses of the two modules tend to stabilize,
and the game's equilibrium $\left(\omega^{*}_L, \omega^{*}_F\right)$ can be obtained.
At equilibrium,
the company generates a new medical image classification model $M_u$ whose parameter $\omega_u$ is $\omega^{*}_F$.
Note that the final $\omega^{*}_L$ and $\omega^{*}_F$ have reached convergence and are similar.
$M_u$ removes the $u$'s data and performs similarly to the retrained model,
with a classification accuracy roughly aligning with $M_o$ (e.g., a $1.5\%$ decrease).
Meanwhile, the model mitigates the privacy leakage risks;
thus,
the attacker's advantage over $D_f$ is very small.



\section{Experiments} \label{experiment}

\subsection{Experimental setup}

The experiments were conducted on a server with Xeon(R) Platinum 8352V 2.10GHz CPU and NVIDIA A100 GPU with 80G RAM.
We applied Adult, MNIST, CIFAR10, and SVHN databases to implement our experiment. 
	
	
	
For text dataset Adult,
we adopted the classic machine learning model,
multilayer perceptron (MLP) with a simple structure.
For the image datasets,
MNIST,
CIFAR$10$ and SVHN,
we used convolutional neural network (CNN) models,
DenseNet, and ResNet$18$.
In addition,
a binary classifier
was employed to train membership inference attack model $M_A$ \cite{chen2021when}.
In a random unlearning scenario,
the unlearning rate is $1$\%,
$2$\%,
$5$\% and $10$\% of the original training set $D_{train}$.
Additionally,
for data removal,
an alternative model $M_r^{'}$ was trained on $D_r^{'}$,
a subset of the retain training set $D_r$ ($D_r^{'} = 20\% \times D_r$).


We used the following metrics to analyze the proposed method:

\textbf{Accuracy.}
We calculated the overall accuracy of the unlearned model $M_u$ and the retrained model $M_r$.

\textbf{Privacy attack advantage.}
We defined a metric called as privacy attack advantage ($PriAA$) as shown in Eq. \ref{priaa} based on membership inference attacks.
A small $PriAA$ value represents lower privacy risks.

\textbf{Running time.}
We investigated the time required to implement the proposed method and made comparisons with retraining from scratch.
	
\textbf{Attack success probability of MIA.}
We employed membership inference attacks \cite{shokri2017} to verify whether the information about $D_f$ was successfully removed from the original model $M_o$.
The attack probability should be lower in the unlearned model $M_u$ for the samples in $D_f$.
	
\textbf{Loss curve.}
We visualized the changes in the proposed method's player loss functions during the iteration process,
thereby demonstrating game convergence.

\subsection{Results and analyses}

\subsubsection{Sample-level unlearning}

We  first conducted experiments on sample-level unlearning,
with $1$\%,
$2$\%,
$5$\%, and $10$\% of the training data deleted randomly.

\paragraph{Accuracy}

The accuracy of the unlearned model $M_u$ is crucial in evaluating the proposed method's classification performance.
Thus,
in this experiment,
we calculated $M_u$'s accuracy for the above four datasets,
comparing the results with those of $M_r$,
as shown in Table. \ref{sampleacc}.

\begin{table}[h] \scriptsize
	\begin{center}
		\caption{The Accuracy of Unlearned Model and Retrained Model at Sample-level Unlearning}
		\label{sampleacc}
		\begin{tabular}{c|c|c|c|c|c}
			\toprule
			\multirow{2}{*}{Dataset} & \multirow{2}{*}{Model} & \multirow{2}{*}{$D_f$} & \multicolumn{3}{c}{Accuracy} \\ \cline{4-6}
			& & & Original Model & Retrained Model & Our Method \\ 
			\midrule
			\multirow{4}{*}{Adult} & \multirow{4}{*}{MLP} & $1\%$ & $84.88$ & $84.18$ & $84.88$ \\
			& & $2\%$ & $84.88$ & $84.67$ & $84.92$ \\ 
			& & $5\%$ & $84.88$ & $84.50$ & $84.89$ \\ 
			& & $10\%$ & $84.88$ & $84.06$ & $84.87$ \\ \cline{1-6}
			
			\multirow{4}{*}{MNIST} & \multirow{4}{*}{Resnet$18$} & $1\%$ & $99.45$ & $99.45$ & $99.27$ \\ 
			& & $2\%$ & $99.45$ & $99.44$ & $99.24$ \\ 
			& & $5\%$ & $99.45$ & $99.47$ & $99.30$ \\ 
			& & $10\%$ & $99.45$ & $99.30$ & $99.29$ \\ \cline{1-6}
			
			\multirow{4}{*}{CIFAR10} & \multirow{4}{*}{Densenet} & $1\%$ & $56.35$ & $49.75$ & $48.66$ \\ 
			& & $2\%$ & $56.35$ & $49.60$ & $48.38$ \\ 
			& & $5\%$ & $56.35$ & $48.25$ & $48.32$ \\ 
			& & $10\%$ & $56.35$ & $48.93$ & $48.32$ \\ \cline{1-6}
			
			\multirow{4}{*}{SVHN} & \multirow{4}{*}{Densenet} & $1\%$ & $90.23$ & $89.76$ & $87.80$ \\
			& & $2\%$ & $90.23$ & $89.64$ & $87.78$ \\
			& & $5\%$ & $90.23$ & $89.85$ & $87.35$ \\ 
			& & $10\%$ & $90.23$ & $88.96$ & $87.28$ \\ 
			\bottomrule
		\end{tabular}
	\end{center}
\end{table}

According to the table,
we observe that the unlearned model generated by our method performs similarly to the retrained model.
In the $1$\% random unlearning on Adult + MLP,
our method obtains an accuracy of $84.88$\% which is slightly higher than that of the retraining model $84.18$\%.
The proposed method's accuracy on MNIST + ResNet$18$ and SVHN + DenseNet was $99.27$\% and $87.80$\% when the  unlearning rate is $1$\%, 
respectively.
This is slightly similar to the retrained model's results.
Similar results can also be obtained when the training samples' unlearning rate is $2$\%, $5$\%, and $10$\%.
On the CIFAR$10$ + DenseNet,
the unlearned model's accuracy is comparatively low, with an average value of $48$\%,
due to the original model's poor performance.
However,
it can be seen that the proposed method's accuracy is quite similar to that of the retrained model.
This indicates that although the initial machine learning process is not ideal,
our method remains effective.
Meanwhile,
as the amount of unlearned data $D_f$ increases,
the proposed method's accuracy declines due to the removal of more information.
Table \ref{sampleacc} illustrates that the model performance of our method is comparable to that of the retrained model.

\paragraph{Privacy risks}

\begin{figure*}[h]
	\centering
	\subfloat[Adult dataset]{
		\includegraphics[scale=0.2]{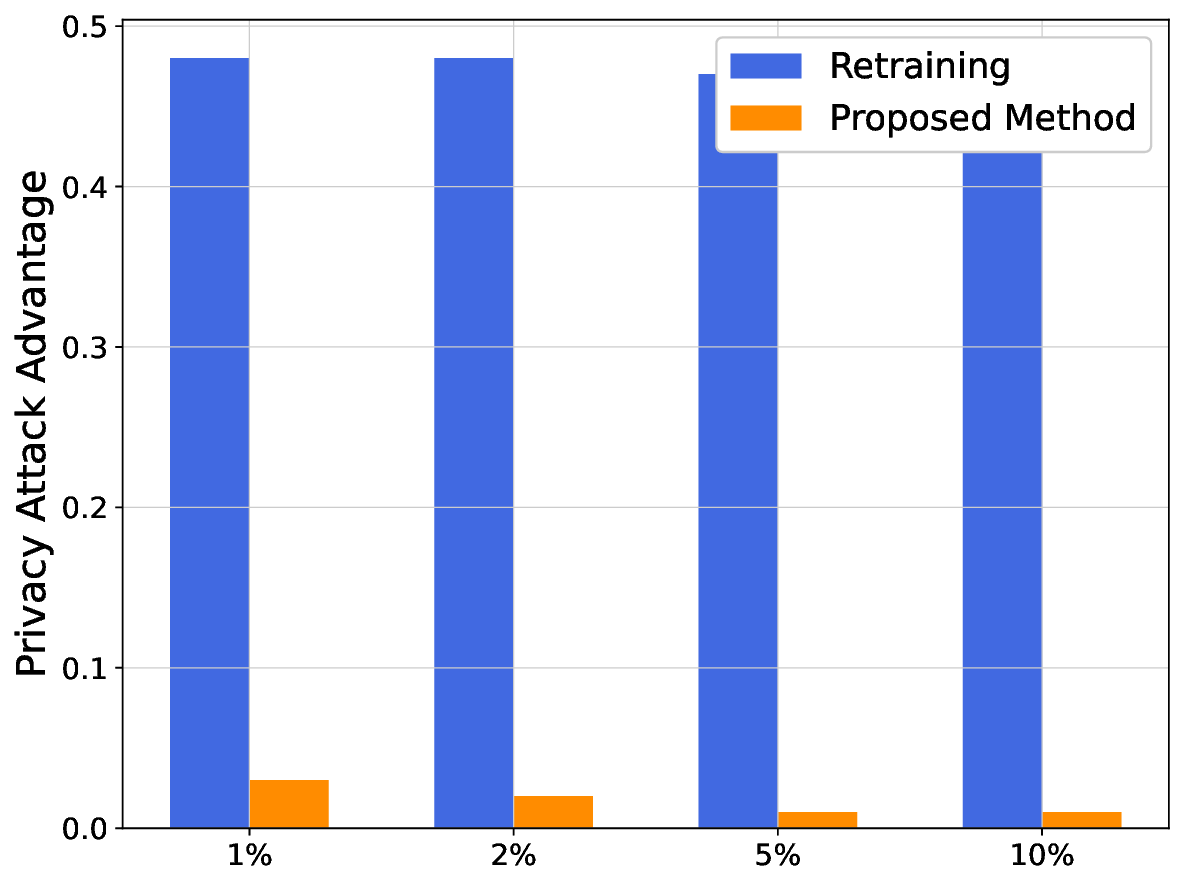}
		\label{sampleprivacy-adult}}
	\subfloat[MNIST dataset]{
		\includegraphics[scale=0.2]{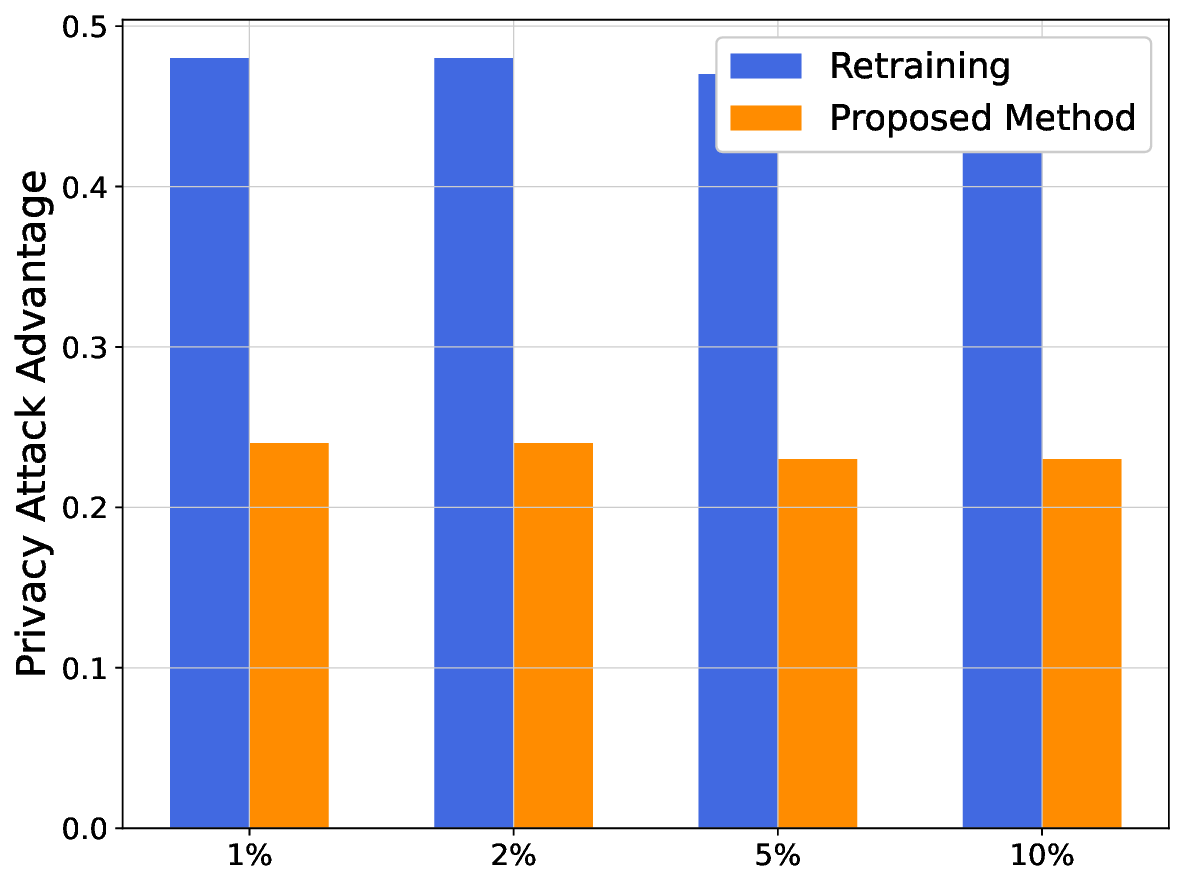}
		\label{sampleprivacy-mnist}}
	\subfloat[CIFAR10 dataset]{
		\includegraphics[scale=0.2]{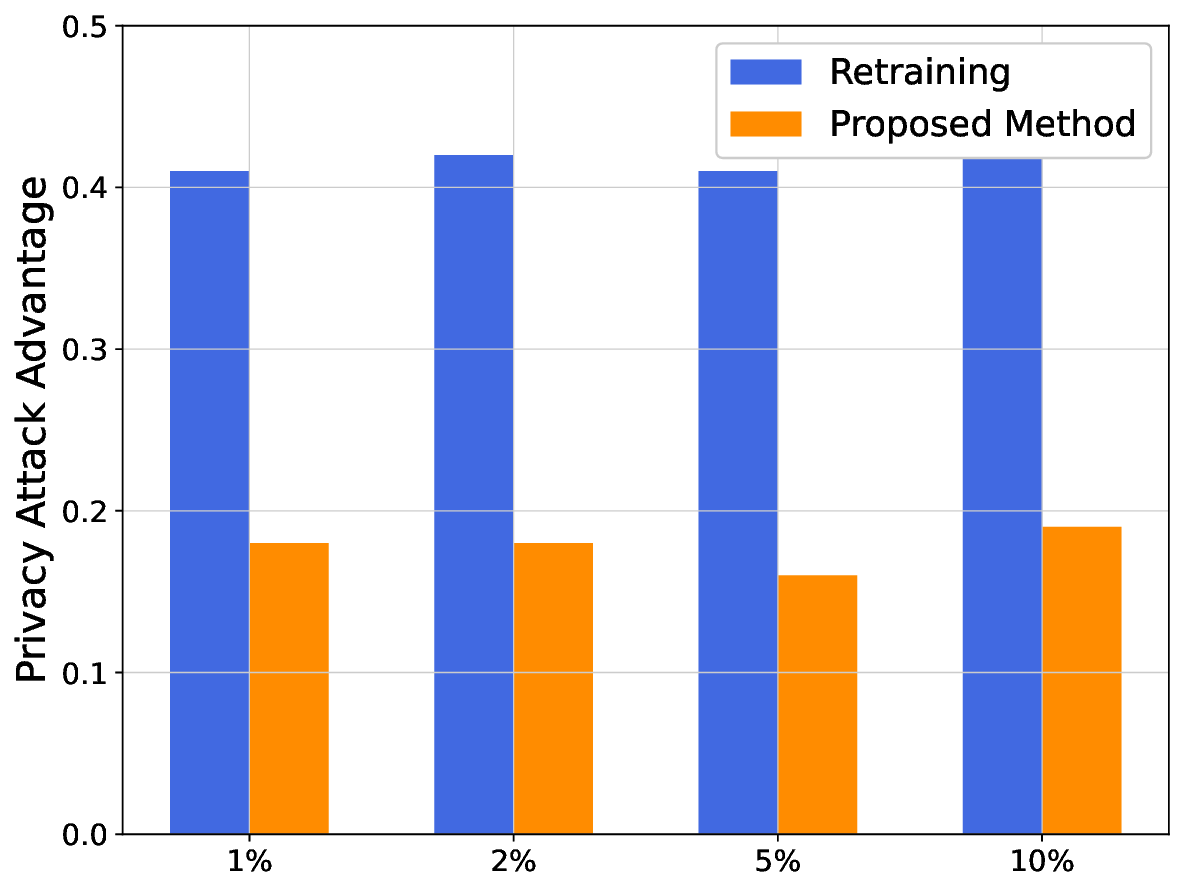}
		\label{sampleprivacy-cifar10}}
	\subfloat[SVHN dataset]{
		\includegraphics[scale=0.2]{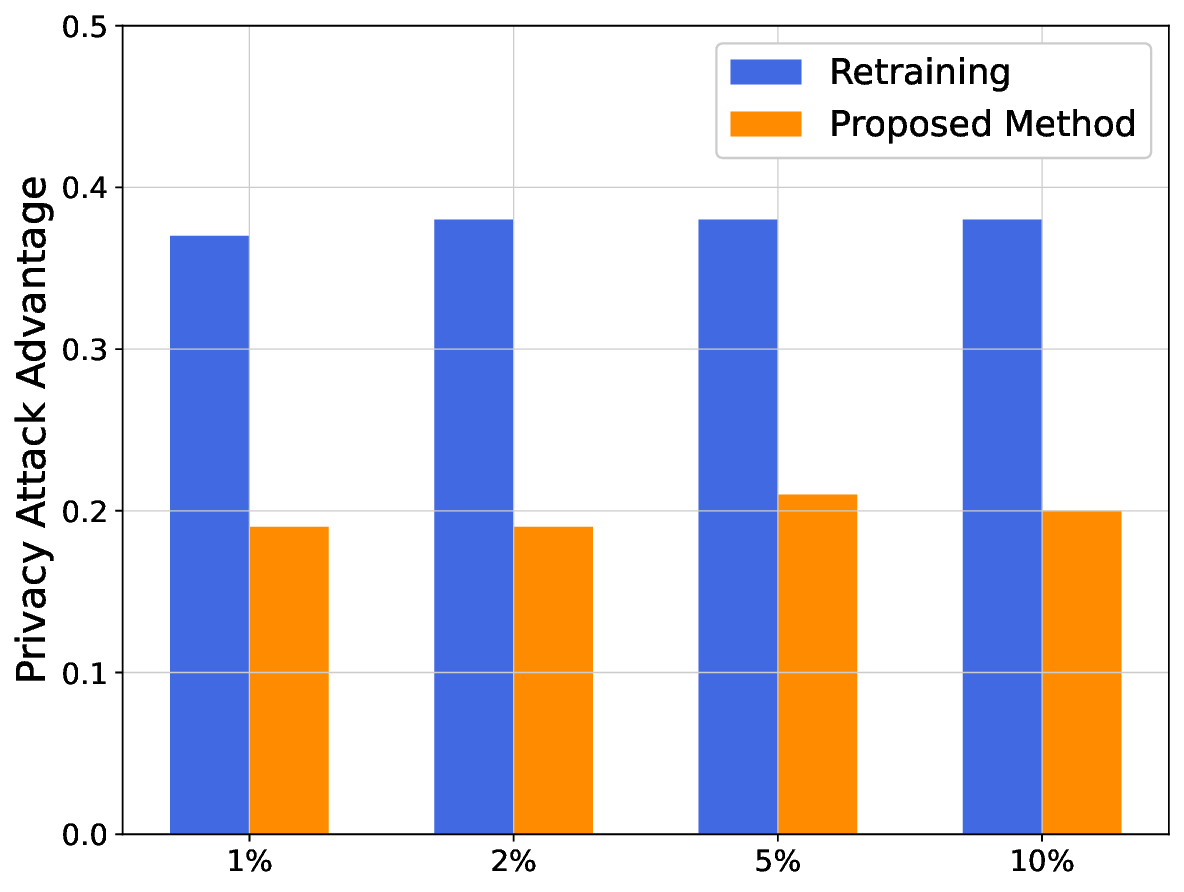}
		\label{sampleprivacy-svhn}}
	
	\caption{The privacy risks of our method compared with retraining at sample-level unlearning}
\end{figure*}

The next experiment examines the proposed method's privacy disclosure risks through training a specific membership inference attack according to the method in \cite{chen2021when}.
This model is used to calculate the privacy attack advantage.
Intuitively, 
the privacy attack advantage's value should be lower for the unlearned model.
Fig. \ref{sampleprivacy-adult},
Fig. \ref{sampleprivacy-mnist},
Fig. \ref{sampleprivacy-cifar10}, and Fig. \ref{sampleprivacy-svhn} show the privacy attack advantage of our method and the model retrained on four datasets,
respectively.

As shown in Fig. \ref{sampleprivacy-adult},
on the Adult dataset,
our method obtains a $D_f$'s privacy confidence value below $0.1$,
which approximates $0$ when removing $10\%$ of training samples,
indicating that the proposed method reduces the risks of membership information leakage.
This is because our method's privacy module selects the best strategy in each game to meet the privacy protection requirements.
Compared to retraining with a value of approximately $0.5$,
the game-theoretical unlearning method significantly improves privacy.
From the results,
retraining enables the attacker to accurately infer the data's membership,
despite achieving performance similar to the original model.
This also confirms that, 
when considering the trade-off between utility and privacy, 
retraining may not be the ideal solution.
Similar patterns can also be observed in Fig. \ref{sampleprivacy-mnist} to Fig. \ref{sampleprivacy-svhn} using different datasets and models.
Our method achieves a confidence advantage of approximately $0.2$ on three image datasets.
This is because feature information in image data easier to extract using membership inference attacks than text data.
However,
the values remain lower than the retrained model's baseline.
The experimental results demonstrate that the proposed method effectively mitigates the privacy risks caused by the vulnerability of unlearning.

\paragraph{Efficiency}

Evaluating the implementation efficiency of an unlearning algorithm is crucial in practical applications.
The model provider should respond to user requests promptly without excessive computational resource consumption.
In this experiment,
we used running time to assess our method's efficiency,
and compared it with the retrained models.
Table \ref{sampletime} displays the results under different datasets and unlearning rates.

\begin{table}[h]
	\begin{center}
		\caption{The Running Time of Our Method Compared with Retraining at Sample-level Unlearning}
		\label{sampletime}
		\begin{tabular}{c|c|c|c|c|c}
			\toprule
			\multirow{2}{*}{Dataset} & \multirow{2}{*}{Model} & \multirow{2}{*}{$D_f$} & \multicolumn{3}{c}{Running Time (s)} \\ \cline{4-6}
			& & & Retraining & Our Method & Speed up \\ 
			\midrule
			\multirow{4}{*}{Adult} & \multirow{4}{*}{MLP} & $1\%$ & $7.8$ & $5.9$ & $1.3\times$ \\
			& & $2\%$ & $7.8$ & $5.4$ & $1.4\times$ \\ 
			& & $5\%$ & $7.9$ & $3.5$ & $2.3\times$ \\ 
			& & $10\%$ & $7.7$ & $3.3$ & $2.3\times$ \\ \cline{1-6}
			
			\multirow{4}{*}{MNIST} & \multirow{4}{*}{Resnet18} & $1\%$ & $2595.8$ & $176.4$ & $14.7\times$ \\
			& & $2\%$ & $2468.3$ & $183.2$ & $13.5\times$ \\ 
			& & $5\%$ & $2339.9$ & $190.6$ & $12.3\times$ \\ 
			& & $10\%$ & $2225.6$ & $203.3$ & $11.0\times$ \\ \cline{1-6}
			
			\multirow{4}{*}{CIFAR10} & \multirow{4}{*}{Densenet} & $1\%$ & $2034.9$ & $54.1$ & $37.6\times$ \\
			& & $2\%$ & $1942.4$ & $60.0$ & $32.4\times$ \\ 
			& & $5\%$ & $1857.4$ & $79.4$ & $23.4\times$ \\ 
			& & $10\%$ & $1720.6$ & $79.5$ & $21.6\times$ \\ \cline{1-6}
			
			\multirow{4}{*}{SVHN} & \multirow{4}{*}{Densenet} & $1\%$ & $1709.3$ & $73.7$ & $23.2\times$ \\
			& & $2\%$ & $1672.8$ & $77.2$ & $21.7\times$ \\ 
			& & $5\%$ & $1574.1$ & $93.5$ & $16.8\times$ \\ 
			& & $10\%$ & $1573.0$ & $119.8$ & $13.1\times$ \\ 
			\bottomrule
		\end{tabular}
	\end{center}
\end{table}

From the table,
it can be seen that retraining is the most time-consuming unlearning strategy.
In contrast,
our method reduces the execution time for data removal.
For example,
on MNIST + ResNet$18$,
the proposed method's running time is at least 10 times faster than retraining,
because only unlearned  and few training samples are required to complete the unlearning process.
Similar results were achieved for the CIFAR$10$ and SVHN datasets,
with an acceleration rate of up to $37$ times faster than retraining.
In comparison,
the efficiency advantage for the Adult dataset is not as pronounced,
with a rate $2$ times faster than retraining.
This is because text data features are easy to learn,
and the model structure is relatively simple.
Notably, 
the time increases with a rise in the unlearning
rate,
as the privacy module needs to calculate payoffs related to $D_f$.
Overall,
in terms of unlearning efficiency,
the proposed method made significant improvements.

\paragraph{Attack success probability of MIA}

The essential goal of machine unlearning is to erase requested data and its influence from the original model to protect privacy.
We estimate the amount of information about unlearned data present within the model using a traditional membership inference attack \cite{shokri2017}.
A lower attack probability indicates a higher data unlearning success rate.
The results are shown in Table \ref{samplemia}.

\begin{table}[h] \scriptsize
	\begin{center}
		\caption{The Attack Success Probability of MIA on Our Method and Retraining at Sample-level Unlearning}
		\label{samplemia}
		\begin{tabular}{c|c|c|c|c|c}
			\toprule
			\multirow{2}{*}{Dataset} & \multirow{2}{*}{Model} & \multirow{2}{*}{$D_f$} & \multicolumn{3}{c}{Attack Success Probability of MIA} \\ \cline{4-6}
			& & & Original Model & Retraining & Our Method \\
			\midrule
			\multirow{4}{*}{Adult} & \multirow{4}{*}{MLP} & $1\%$ & $1.0$ & $0.48$ & $0.00$  \\
			& & $2\%$ & $1.0$ & $0.48$ &  $0.03$ \\ 
			& & $5\%$ & $1.0$ & $0.62$ & $0.01$  \\ 
			& & $10\%$ & $1.0$ & $0.60$ &  $0.00$ \\  \cline{1-6}
			
			\multirow{4}{*}{MNIST} & \multirow{4}{*}{Resnet18} & $1\%$ & $1.0$ & $0.57$ &  $0.00$ \\
			& & $2\%$ & $1.0$ & $0.56$ &  $0.00$ \\ 
			& & $5\%$ & $0.99$ & $0.57$ & $0.00$  \\ 
			& & $10\%$ & $0.97$ & $0.55$ & $0.00$  \\  \cline{1-6}
			
			\multirow{4}{*}{CIFAR10} & \multirow{4}{*}{Densenet} & $1\%$ & $1.0$ & $0.51$ & $0.04$  \\
			& & $2\%$ & $1.0$ & $0.51$ & $0.04$  \\ 
			& & $5\%$ & $0.99$ & $0.56$ &  $0.03$ \\ 
			& & $10\%$ & $0.99$ & $0.52$ & $0.03$ \\  \cline{1-6}
			
			\multirow{4}{*}{SVHN} & \multirow{4}{*}{Densenet} & $1\%$ & $0.99$ & $0.55$ &  $0.00$ \\
			& & $2\%$ & $0.99$ & $0.54$ &  $0.00$ \\ 
			& & $5\%$ & $0.98$ & $0.49$ &  $0.00$ \\ 
			& & $10\%$ & $0.98$ & $0.49$ &  $0.00$ \\
			\bottomrule
		\end{tabular}
	\end{center}
\end{table}

Table \ref{samplemia} exhibits the membership inference attack results.
The membership inference attacks results on the original model are nearly $100$\%,
meaning that the attacker can successfully infer whether the target sample belongs to the original training set.
Comparing the fourth and sixth columns in Table \ref{samplemia} shows that the attack probability drops significantly,
which indicates that the attacker could not determine if the target data was used during the training process. Specifically,
the proposed method successfully removed the requested data and its influence on the model.
Furthermore, 
our method's unlearned model attack probability was lower than that of retraining from scratch,
implying that it contains less information.
This experiment verifies that the proposed unlearning method can indeed erase the unlearned data from a trained model.

\paragraph{Convergence}

\begin{figure}[h]
	\centering
	\subfloat[Adult dataset ($1\%$)]{
		\includegraphics[scale=0.3]{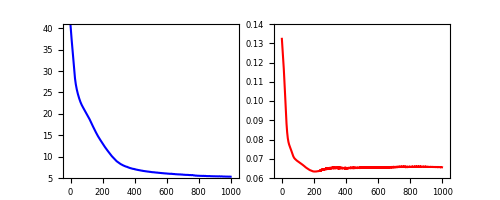}}
	\subfloat[Adult dataset ($2\%$)]{
		\includegraphics[scale=0.3]{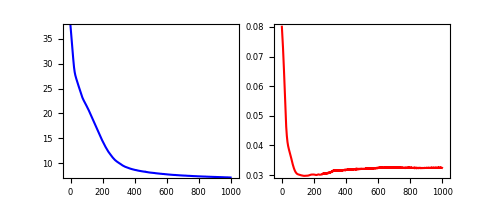}}
	\\
	\subfloat[Adult dataset ($5\%$)]{
		\includegraphics[scale=0.3]{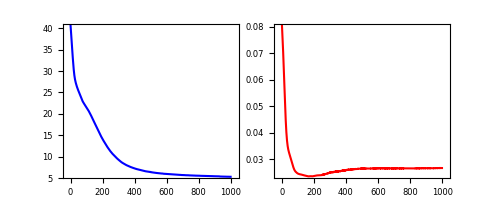}}
	\subfloat[Adult dataset ($10\%$)]{
		\includegraphics[scale=0.3]{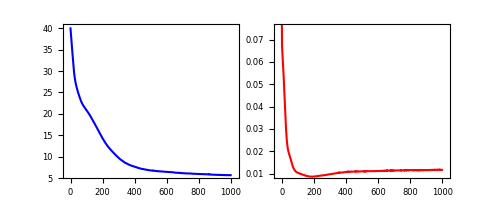}}
	\\
	\subfloat[MNIST dataset ($1\%$)]{
		\includegraphics[scale=0.3]{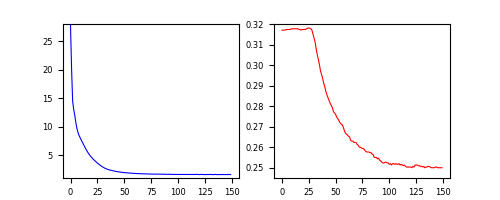}}
	\subfloat[MNIST dataset ($2\%$)]{
		\includegraphics[scale=0.3]{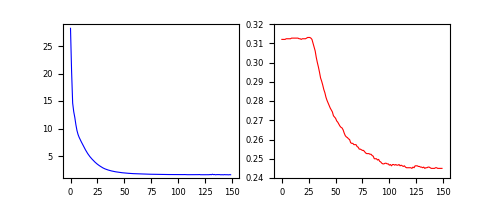}}
	\\
	\subfloat[MNIST dataset ($5\%$)]{
		\includegraphics[scale=0.3]{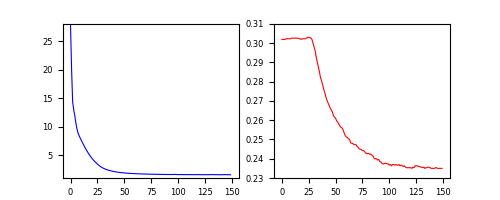}}
	\subfloat[MNIST dataset ($10\%$)]{
		\includegraphics[scale=0.3]{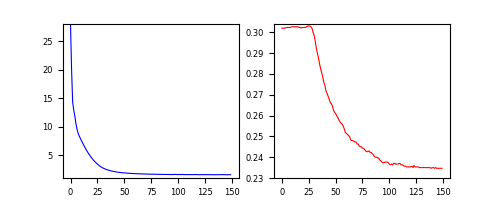}}
	\\
	\subfloat[CIFAR10 dataset ($1\%$)]{
		\includegraphics[scale=0.3]{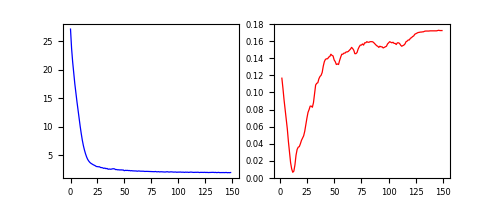}}
	\subfloat[CIFAR10 dataset ($2\%$)]{
		\includegraphics[scale=0.3]{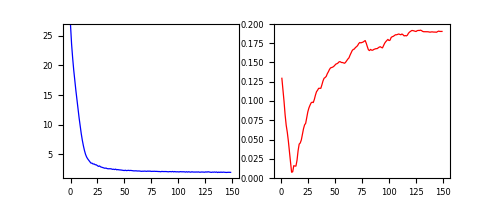}}
	\\
	\subfloat[CIFAR10 dataset ($5\%$)]{
		\includegraphics[scale=0.3]{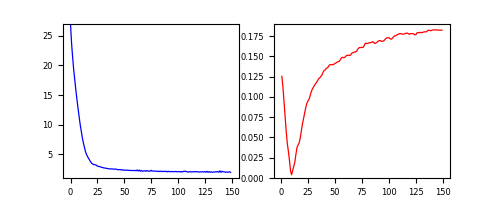}}
	\subfloat[CIFAR10 dataset ($10\%$)]{
		\includegraphics[scale=0.3]{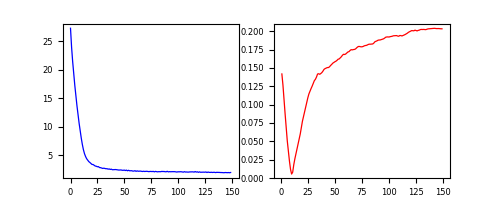}}
	\\
	\subfloat[SVHN dataset ($1\%$)]{
		\includegraphics[scale=0.3]{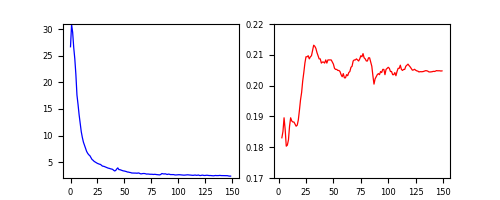}}
	\subfloat[SVHN dataset ($2\%$)]{
		\includegraphics[scale=0.3]{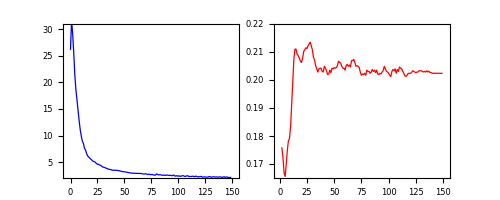}}
	\\
	\subfloat[SVHN dataset ($5\%$)]{
		\includegraphics[scale=0.3]{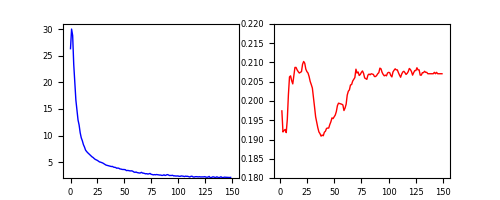}}
	\subfloat[SVHN dataset ($10\%$)]{
		\includegraphics[scale=0.3]{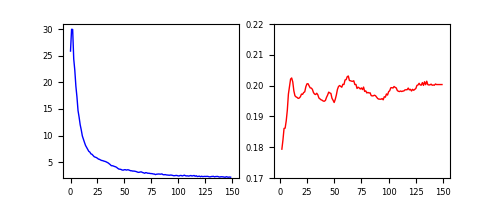}}
	
	\caption{The loss of our method at sample-level unlearning}
	\label{sample-c}
\end{figure}

Current works on machine unlearning have not focused on the potential of method convergence.
Since our method is based on the game theory,
we ensured its effectiveness by visualizing the changes of loss functions of the unlearning module and privacy module during the iteration process. Fig. \ref{sample-c} displays the convergence results.

According to the figure, 
after completing all the games,
the players' loss functions converge into a stable state.
For example,
on Adult + MLP,
the unlearning and privacy modules' losses directly decrease to the lowest point and then stabilize,
indicating that both players have found the optimal equilibrium solution. 
Specifically,
the resulting unlearned model achieves a relative balance between performance and privacy protection.
A similar trend can also be observed in Fig. \ref{sample-c} on the MNIST dataset,
but the privacy module's loss remains unchanged for a short period before declining.
This may be because,
at first,
fine-tuning of the unlearning module did not significantly reduce the information amount of the target samples.
Conversely,
the losses of the privacy module in Fig. \ref{sample-c} initially decrease, then increase, and finally stabilize, indicating that the information changes contained in the intermediate parameters can be extracted by the attacker.
This experiment validates the proposed method's feasibility.

\subsubsection{Class-level unlearning}

In real-world scenarios,
unlearning involves deleting single or several records with different labels.
Users can also request the removal of all the samples belong to a single class.
A machine unlearning method should be general,
Ideally,
it should be able to solve sample and class-level unlearning.
Thus,
we examined the proposed method's effectiveness in erasing class-specific data.

First,
we evaluated the proposed method's performance in class-level unlearning.
Unlike the evaluation for random samples removal,
we calculated the model's classification performance on the unlearned and retained classes.
Table \ref{classacc} shows the accuracy of the original model,
the retrained model and the unlearned model obtained by our method on $D_f$ and $D_r$,
respectively.

\begin{table}[h]
	\begin{center}
		\caption{The Accuracy of Unlearned Model and Retrained Model at Class-level Unlearning}
		\label{classacc}
		\begin{tabular}{c|cc|cc|cc}
			\toprule
			 \multirow{3}{*}{Class} & \multicolumn{6}{c}{Accuracy} \\ \cline{2-7}
			& \multicolumn{2}{c}{Original Model} & \multicolumn{2}{c}{Retrained Model} & \multicolumn{2}{c}{Our Method} \\ \cline{2-7}
			 & $D_f$ & $D_r$ & $D_f$ & $D_r$ & $D_f$ & $D_r$ \\
			\midrule
			 0 airplane & $63.96$ &	$55.48$ &	$0.00$ &	$49.63$ &	$13.43$ &	$48.93$\\
			 1 automobile &  $65.83$ &	$55.25$ &	$0.00$ &	$49.39$ &	$5.62$ &	$46.43$\\ 
			2 bird & $42.60$ &	$57.86$ &	$0.00$ &	$52.62$ &	$7.81$ &	$50.40$\\ 
			3 cat &  $38.85$ &	$58.19$ &	$0.00$ &	$52.83$ &	$9.06$ &	$51.37$\\
			 4 deer & $45.52$ &	$57.52$ &	$0.00$&	$51.57$ &	$9.58$ &	$52.08$ \\  
			 5 dog & $46.98$ &	$57.44$ &	$0.00$ &	$51.75$ &	$7.60$ &	$49.31$ \\  
			 6 frog & $64.17$ &	$55.49$ &	$0.00$ &	$48.38$ &	$3.54$ &	$48.09$ \\  
			 7 horse & $61.25$ &	$55.83$ &	$0.00$ &	$50.11$ &	$2.50$ &	$48.31$ \\ 
			 8 ship & $69.68$ &	$54.75$ &	$0.00$ &	$49.95$ &	$5.41$ &	$47.92$ \\  
			 9 truck &  $64.47$ &	$55.35$ &	$0.00$ &	$50.26$ &	$5.52$ &	$46.13$\\ 
			\bottomrule
		\end{tabular}
	\end{center}
\end{table}

Based on the table,
we can observe that compared to the original model,
the proposed method achieves a lower accuracy on the unlearned class,
indicating that our method removes a class's training samples and their accompanying information.
Taking the eighth class,
"horse",
as an example,
our method obtains an accuracy of $48.31\%$ on the retained classes, 
similar to the retained model's $50.11\%$.
The accuracy on the unlearned class also decreases from the original model’s $61.25\%$ to $2.5\%$.
This is inconsistent with the retrained model's accuracy,
because the privacy risk generated from the differences between the original and unlearned models were considered in our method.
Similar results can be observed on the remaining nine classes.

\begin{figure}[h]
	\centering
	\includegraphics[scale=0.33]{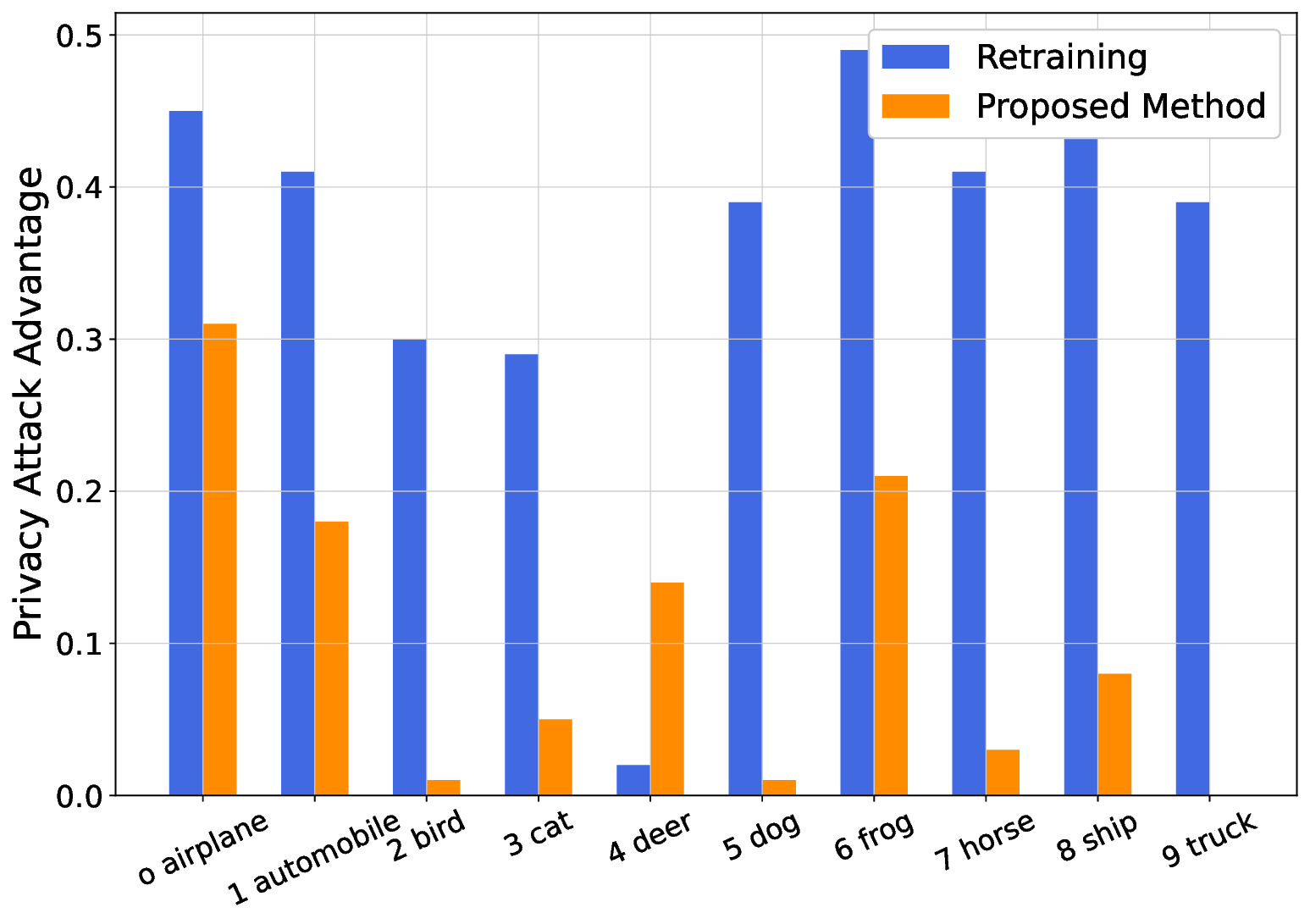}
	\caption{The privacy risks of our method compared with retraining at class-level unlearning}
	\label{classprivacy}
\end{figure}

Subsequently,
we evaluated the proposed method from three aspects:
privacy risk, 
efficiency, 
and membership inference attack success probability.
Fig. \ref{classprivacy} illustrates our method's privacy attack advantages.
It can be seen that,
compared to retraining,
our method significantly reduces the values of privacy attack advantage.
For example,
while unlearning the "truck" class's data, the retraining method generates an unlearned model whose privacy attack advantage is approximately $0.4$.
Specifically,
the attacker can infer the membership of samples.
In contrast,
the proposed game-theoretic method reduces this value to approximately $0$.
This indicates that our method effectively mitigates the potential privacy risks generated during data removal for sample-level and class-level unlearning.

Table \ref{classtime} displays the running time of retraining and our method.
From the table,
we can clearly observe that,
for each class in the CIFAR$10$ dataset,
the proposed method saves nearly $30$ times as much running time compared to retraining,
as it only requires fine-tuning the model parameters using partial data.
This improves the unlearning process's efficiency,
even when tackling the large amounts of data and models with complex structures.

Regarding class-level unlearning,
we employed membership inference attacks for evaluation,
and Table \ref{classmia} shows its attack success probability.
For example, 
the first unlearned class achieved an attack probability of $0.99$ for the original model,
indicating that the unlearned class's membership can  be inferred easily.
For the retrained model,
this value reduces to $0.55$,
while the unlearned model's attack probability obtained by our method was approximately close to $0$,
indicating that the target samples have been successfully removed.
The remaining nine classes obtained similar results.
The experimental results demonstrates that our method can effectively erase the influence of the unlearned class on the original model.

\begin{table}[h]
	\begin{center}
		\caption{The Running Time of Our Method Compared with Retraining at Class-level Unlearning}
		\label{classtime}
		\begin{tabular}{c|c|c|c}
			\toprule
		    \multirow{2}{*}{$D_f$} & \multicolumn{3}{c}{Running Time (s)} \\ \cline{2-4}
			& Retraining & Our Method & Speed up \\
			\midrule
			 0 airplane & $1690.1$ & $58.0$ & $29.1\times$ \\
			 1 automobile & $1701.7$ &  $57.4$ & $29.6\times$ \\ 
			 2 bird & $1706.3$ & $57.8$  & $29.5\times$ \\ 
		   	 3 cat & $1825.5$ &  $57.3$ & $31.8\times$ \\
			 4 deer & $1774.8$ &  $57.1$ & $31.1\times$ \\  
			 5 dog & $1858.1$ &  $56.8$ & $32.7\times$ \\  
		     6 frog & $1889.3$ &  $63.5$ & $29.7\times$ \\  
			 7 horse & $1852.1$ &  $64.0$ & $28.9\times$ \\ 
			 8 ship & $1856.2$ &  $57.7$ & $32.1\times$ \\  
			 9 truck & $1839.7$ &  $58.3$ & $31.5\times$ \\ 
			\bottomrule
		\end{tabular}
	\end{center}
\end{table}

\begin{table}[H]
	\begin{center}
		\caption{The Attack Success Probability of MIA on Our Method and Retraining at Class-level Unlearning}
		\label{classmia}
		\begin{tabular}{c|c|c|c}
			\toprule
		    \multirow{2}{*}{$D_f$} & \multicolumn{3}{c}{Attack Success Probability of MIA} \\ \cline{2-4}
			 & Original Model & Retrained Model & Our Method \\
			\midrule
			0 airplane & $0.99$ & $0.55$ & $0.00$  \\
			1 automobile & $0.99$ & $0.61$ &  $0.00$ \\ 
			2 bird & $0.99$ & $0.47$ & $0.00$  \\ 
			3 cat & $0.98$ & $0.48$ &  $0.00$ \\  
			4 deer & $0.99$ & $0.45$ &  $0.00$ \\  
			5 dog & $0.98$ & $0.55$ &  $0.00$ \\  
			6 frog & $0.99$ & $0.46$ &  $0.00$ \\  
			7 horse & $0.99$ & $0.48$ &  $0.00$ \\  
			8 ship & $0.99$ & $0.58$ &  $0.00$ \\  
			9 truck & $0.98$ & $0.60$ &  $0.00$ \\  
			\bottomrule
		\end{tabular}
	\end{center}
\end{table}

\begin{figure}[h]
	\centering
	\subfloat[Class 0]{
		\includegraphics[scale=0.3]{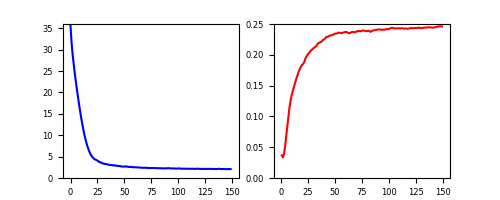}}
	\subfloat[Class 1]{
		\includegraphics[scale=0.3]{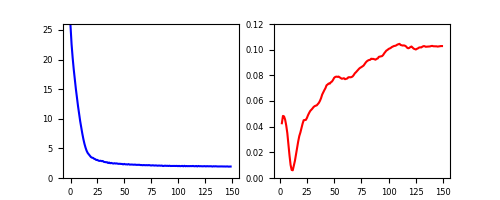}}
	\\
	\subfloat[Class 2]{
		\includegraphics[scale=0.3]{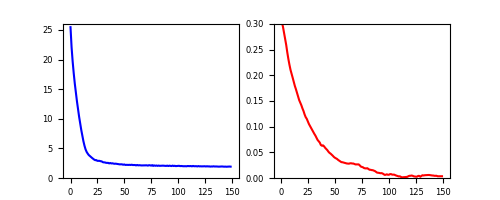}
	\label{class2-c}}
	\subfloat[Class 3]{
		\includegraphics[scale=0.3]{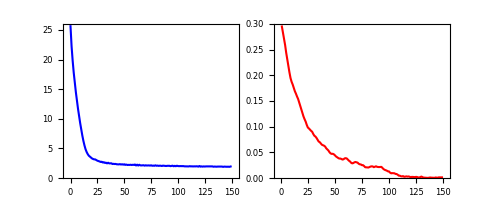}}
	\\
	\subfloat[Class 4]{
		\includegraphics[scale=0.3]{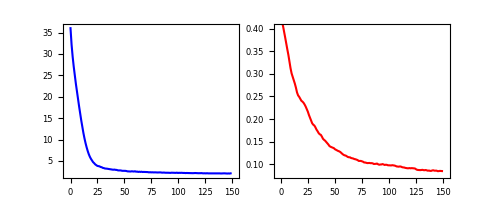}}
	\subfloat[Class 5]{
		\includegraphics[scale=0.3]{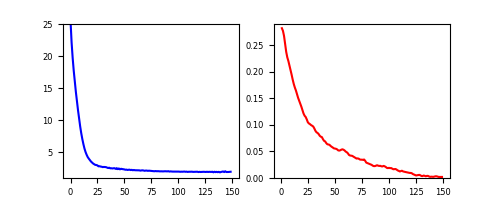}}
	\\
	\subfloat[Class 6]{
		\includegraphics[scale=0.3]{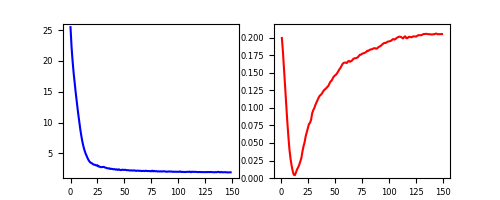}}
	\subfloat[Class 7]{
		\includegraphics[scale=0.3]{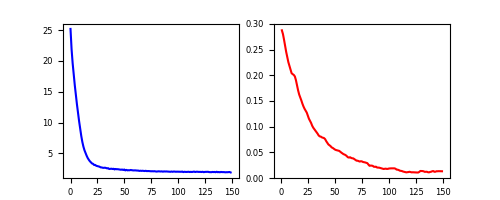}}
	\\
	\subfloat[Class 8]{
		\includegraphics[scale=0.3]{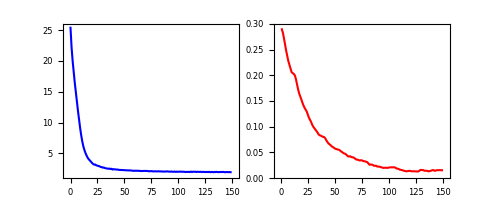}}
	\subfloat[Class 9]{
		\includegraphics[scale=0.3]{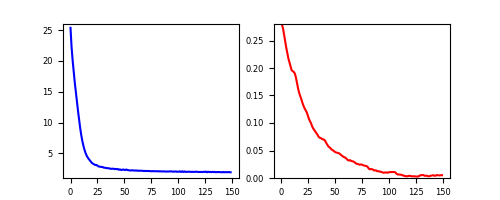}}
	
	\caption{The loss of our method at class-level unlearning}
	\label{cifar10-class-c}
\end{figure}

Finally,
we analyzed the convergence of the proposed method in class-level unlearning through visualizing the loss curves.
Fig. \ref{cifar10-class-c} presents the experimental results for each class on the CIFAR$10$ dataset.
As shown in Fig. \ref{cifar10-class-c},
after completing all the games, 
the players' losses for each unlearned class become relatively stable,
indicating that the game has converged to an equilibrium state and no further strategy changes will occur.
For instance,
in the unlearning process of Class 2 (Fig. \ref{class2-c}),
the losses of the unlearning module and privacy module gradually decrease and stabilize,
indicating that an unlearned model has achieved a balance between utility and privacy.
Although the direction of loss in the privacy module is different due to changes in the information amount contained in the model during the game, 
the overall trend remains stable,
thereby validating our game-theoretic unlearning method's effectiveness.

\subsubsection{Parameter analysis}

\begin{figure}[h]
	\centering
	\subfloat[Adult dataset (1\%)]{
		\includegraphics[scale=0.25]{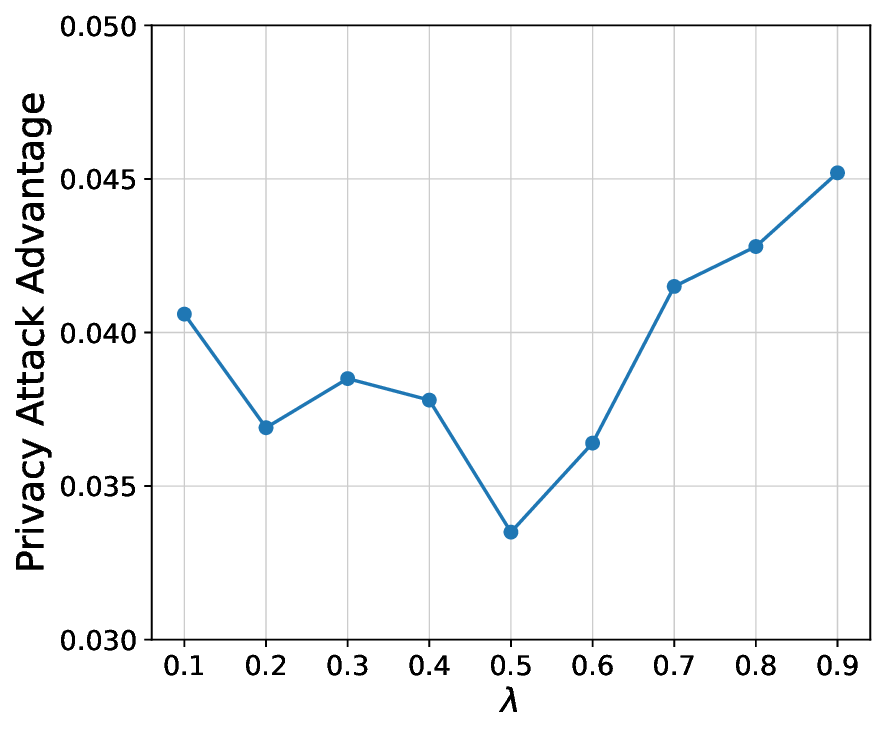}
		\label{adult-p-1}}
	\subfloat[Adult dataset (2\%)]{
		\includegraphics[scale=0.25]{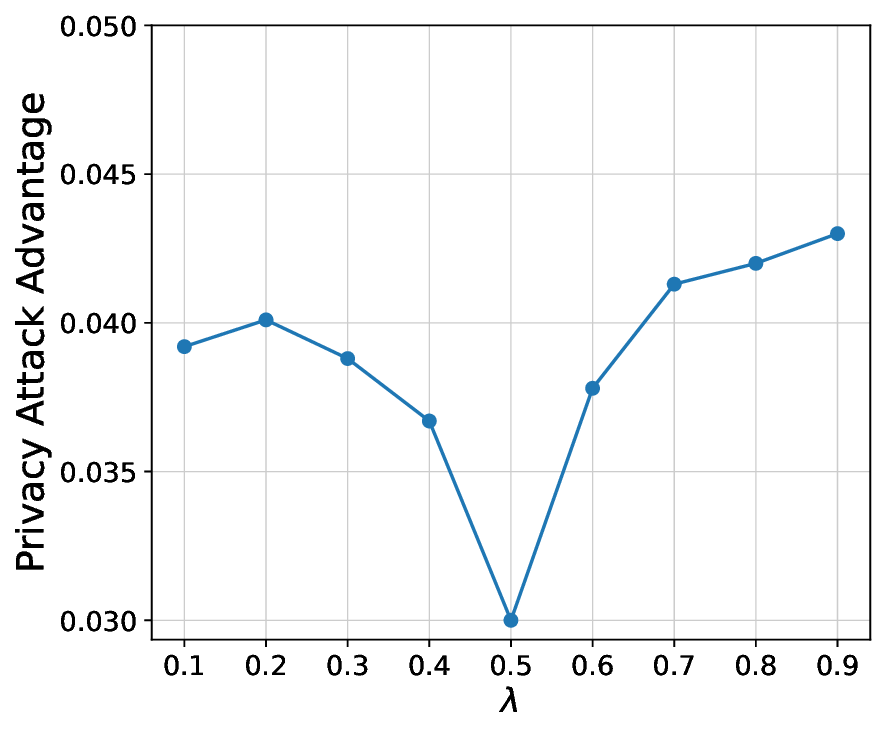}
		\label{adult-p-2}}
	\\
	\subfloat[Adult dataset (5\%)]{
		\includegraphics[scale=0.25]{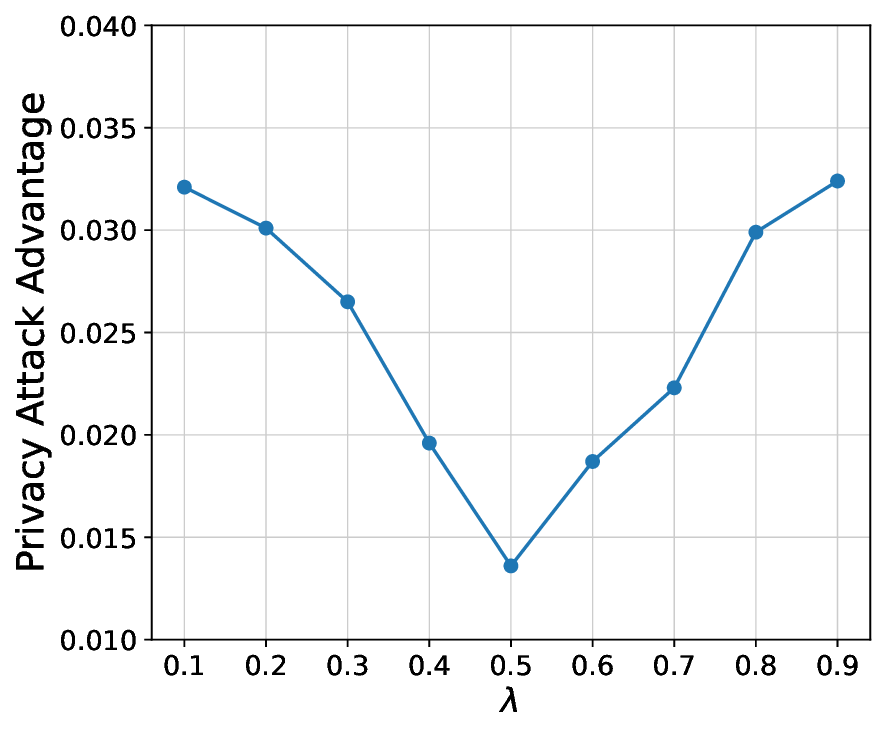}
		\label{adult-p-5}}
	\subfloat[Adult dataset (10\%)]{
		\includegraphics[scale=0.25]{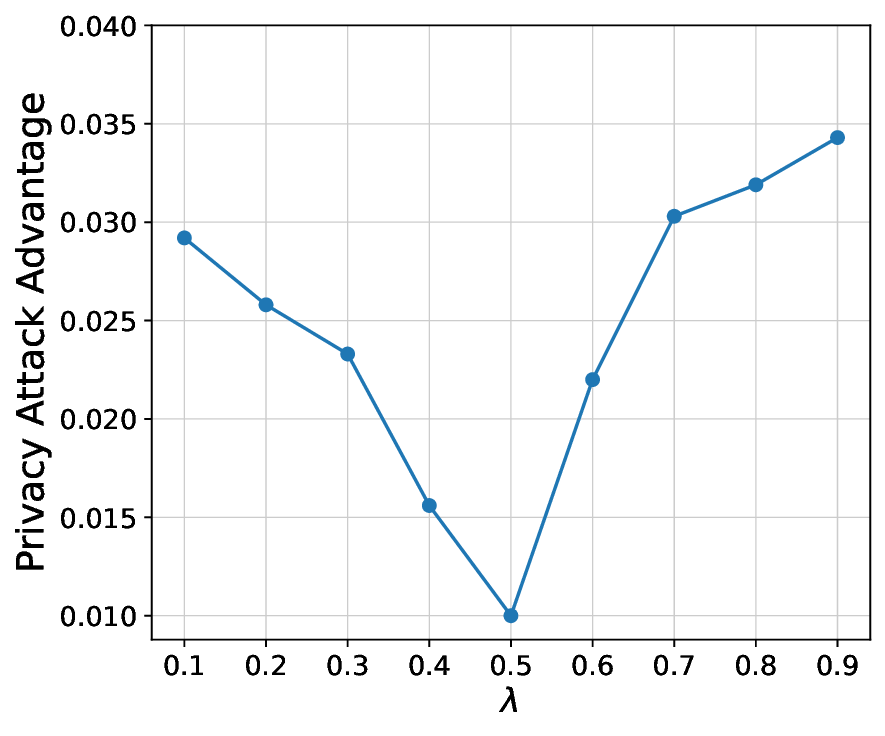}
		\label{adult-p-10}}
	\\
	\subfloat[CIFAR10 dataset (1\%)]{
		\includegraphics[scale=0.25]{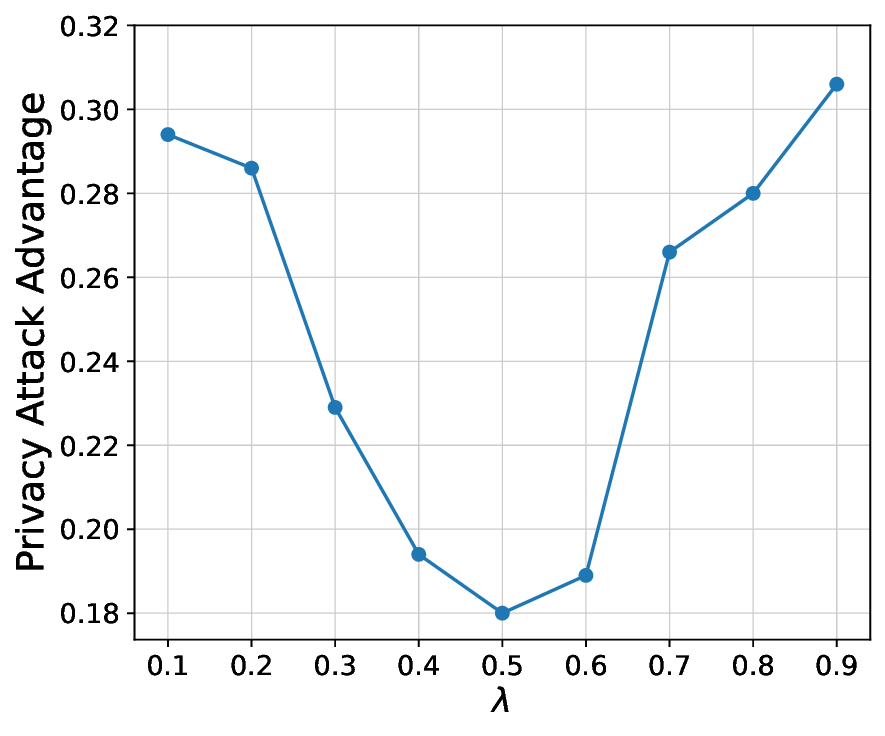}
		\label{cifar10-p-1}}
	\subfloat[CIFAR10 dataset (2\%)]{
		\includegraphics[scale=0.25]{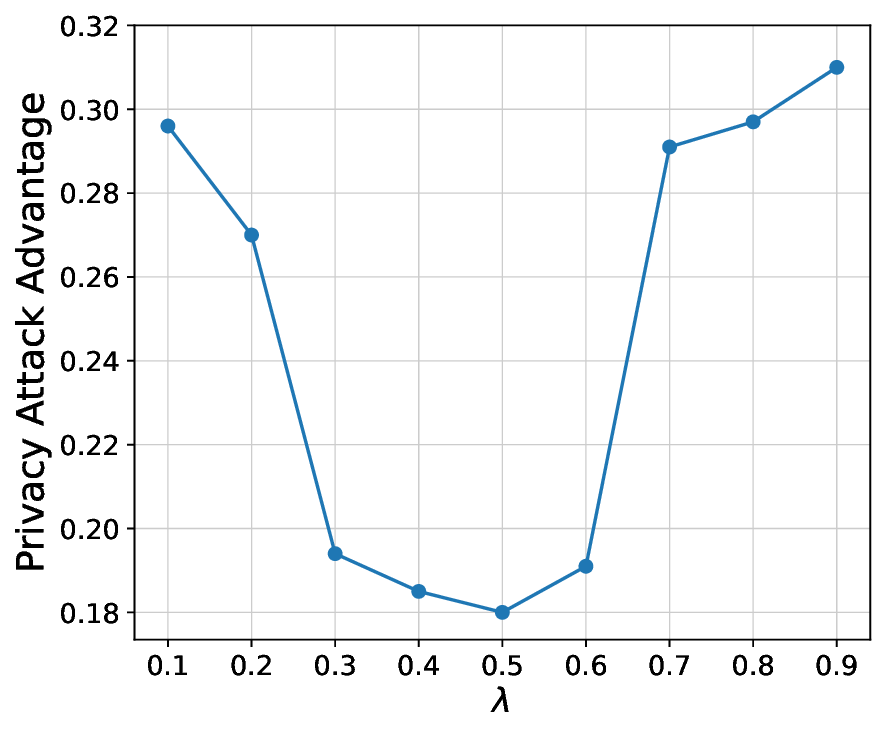}
		\label{cifar10-p-2}}
	\\
	\subfloat[CIFAR10 dataset (5\%)]{
		\includegraphics[scale=0.25]{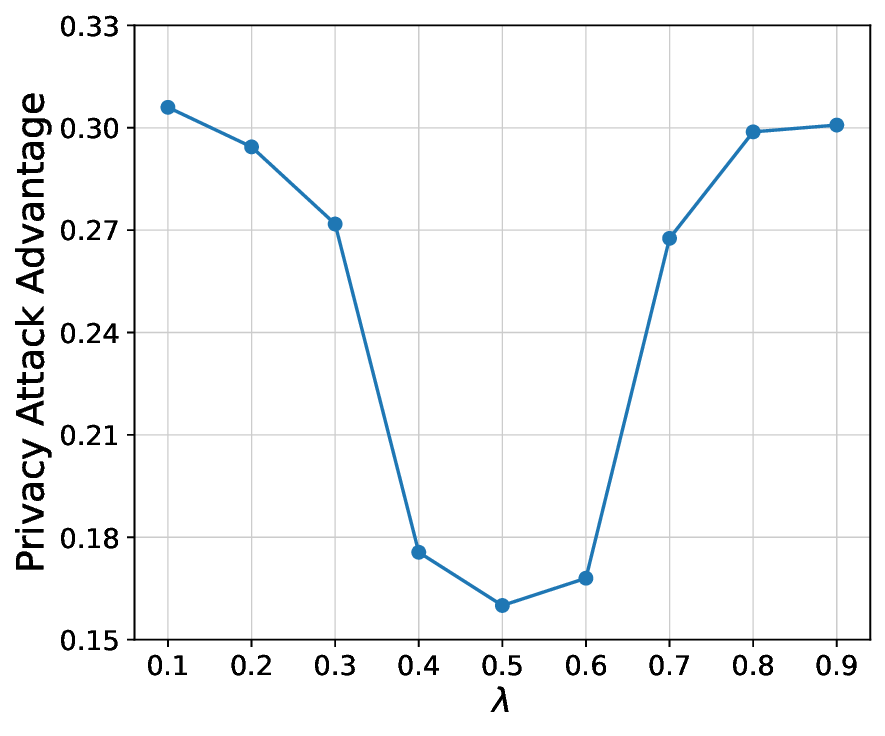}
		\label{cifar10-p-5}}
	\subfloat[CIFAR10 dataset (10\%)]{
		\includegraphics[scale=0.25]{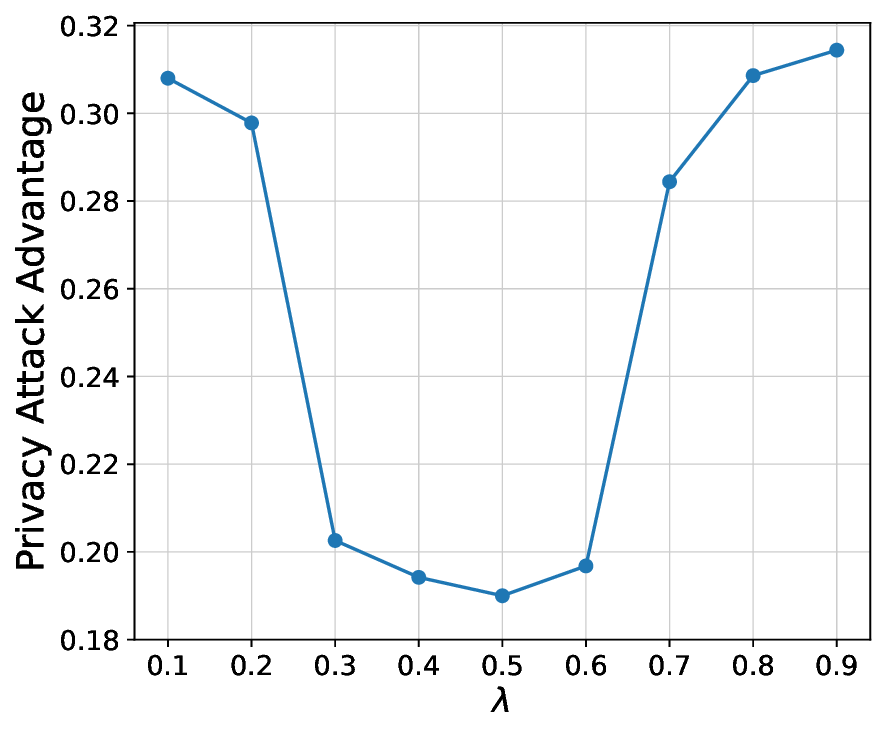}
		\label{cifar10-p-10}}
	
	\caption{The privacy risks of our method with different $\lambda$}
	\label{paramters}
\end{figure}

Our method's privacy module uses $\lambda$ to control the unlearned model's privacy leakage risks during the game process.
Ideally,
$\lambda$ is expected to equal $0.5$,
meaning that it is difficult for the attacker to distinguish between the outputs of the original and unlearned models.
The closer the value of $\lambda$ is to $0$ or $1$,
the less the privacy module moves in the game play,
making the attack probability more significant.
Thus,
we designed an experiment to investigate the influence of $\lambda$ on the unlearned model's privacy risk.
The value of $\lambda$ was set within the range of $0.1 \sim 0.9$,
and the results were assessed by randomly deleting $1\%$,
$2\%$,
$5\%$, and $10\%$ of the training samples.
Fig. \ref{paramters} displays the change in privacy attack advantage with different $\lambda$ values on the Adult and CIFAR$10$ datasets.

As shown in Fig. \ref{adult-p-1} to Fig. \ref{adult-p-10},
it can be observed that as the value of $\lambda$ increases from $0.1$ to $0.9$,
the unlearned model's privacy attack advantage first declines and then rises.
Using $\lambda = 0.5$ as a reference,
the more $\lambda$ deviates from $0.5$,
the higher the privacy attack advantage becomes.
This indicates that the privacy module has lower requirements for privacy protection in the game.
Specifically,
the probability of successful inference by the attacker is higher.
Therefore,
at $\lambda = 0.5$,
the privacy attack advantage is minimized,
and an deviation from this $\lambda$,
increase or decrease in the lambda value raises the risk.
The CIFAR$10$ dataset exhibits similar results in Fig. \ref{cifar10-p-1} to Fig. \ref{cifar10-p-10}.
According to the figures,
we can observe that the unlearned model's privacy attack advantage changes with $\lambda$ due to the privacy module actions during the game,
verifying our method's effectiveness in reducing privacy risks,
and causing the attack probability to approach that of random guessing.

\section{Conclusions} \label{conclusions}

In this paper,
we aim to establish a trade-off between utility and privacy in machine unlearning via modelling,
and we developed a game-theoretic machine unlearning algorithm.
The proposed algorithm comprised an unlearning module and a privacy module.
The former is to ensure the effectiveness of machine unlearning and maintain excellent model performance using an alternative model and an error term.
The latter provides privacy guarantees using a confidence advantage mechanism. 
After both modules find strategies that maximize payoffs from an equilibrium, 
we utilize the final model parameters to generate the unlearned model,
achieving a relative balance between utility and privacy.
The experimental results demonstrate that our method satisfies the data removal requirements,
and performs similarly to the retrained model while effectively mitigating the privacy risks.


\bibliographystyle{IEEEtran}
\bibliography{IEEEabrv,references}

%
\newpage
%
%
%
%
%

\vfill

\end{document}